\documentclass[11pt,a4paper,DIV=13,numbers=noenddot,abstract=true,sfdefaults=false]{scrartcl}

\newcommand{\mytitle}{Application of the 3-Loop FlexibleEFTHiggs Method to the MSSM and the NMSSM}
\newcommand{\myauthors}{Thomas Kwasnitza, Dominik Stöckinger, Alexander Voigt, Johannes Wünsche}
\newcommand{\mykeywords}{Higgs boson mass, supersymmetry}

\usepackage[utf8]{inputenc}
\usepackage[T1]{fontenc}
\usepackage{lmodern}
\usepackage{microtype}
\usepackage{amsmath,amssymb}
\usepackage[sorting=none,citestyle=numeric-comp,eprint=false,isbn=false]{biblatex}
\usepackage{booktabs}
\usepackage{listings}
\usepackage[usenames]{xcolor}\definecolor{fscolor}{RGB}{44,118,255}
\usepackage{xspace}
\usepackage{subcaption}
\usepackage{tikz}
\usepackage{authblk}
\usepackage[format=plain,labelfont={bf}]{caption}
\usepackage{hyperref}
\usepackage{orcidlink} 

\title{\mytitle}

\hypersetup{
  pdftitle    = {\mytitle},
  pdfauthor   = {\myauthors},
  pdfkeywords = {\mykeywords},
  colorlinks  = true,
  linkcolor   = blue,
  citecolor   = blue,
  urlcolor    = blue,
  linkcolor   = blue
}

\usetikzlibrary{calc}

\tikzset{
  scalar/.style={dashed}
}

\newcommand{\abbrev}[1]{\text{#1}} 
\newcommand{\as}{\ensuremath{\alpha_{\text{s}}}}
\newcommand{\aem}{\ensuremath{\alpha_{\text{em}}}}
\newcommand{\BSM}{\abbrev{BSM}\xspace}
\newcommand{\DRbar}{\overline{\abbrev{DR}}\xspace}
\newcommand{\DRbarPrime}{\DRbar'\xspace}
\newcommand{\DeltalambdaSM}{\Delta\lambda_{\SM}}
\newcommand{\EFT}{\abbrev{EFT}\xspace}
\newcommand{\EWSB}{\abbrev{EWSB}\xspace}
\newcommand{\FEFTH}{\texttt{Flex\-ib\-le\-EFT\-Higgs}\@\xspace}
\newcommand{\FH}{\texttt{Feyn\-Higgs}\@\xspace}
\newcommand{\FO}{\abbrev{FO}\xspace}
\newcommand{\FS}{\texttt{Flex\-ib\-le\-SUSY}\@\xspace}
\newcommand{\hc}{\text{h.c.}\xspace}
\newcommand{\Himalaya}{\texttt{Himalaya}\xspace}
\newcommand{\HS}{\abbrev{HS}\xspace}
\newcommand{\HSSUSY}{\texttt{HSSUSY}\xspace}
\newcommand{\lambdaNMSSM}{\lambda}
\newcommand{\lambdaSM}{\hat{\lambda}_{\SM}}
\newcommand{\LL}{\abbrev{LL}}

\newcommand{\LS}{\abbrev{LS}\xspace}
\newcommand{\MDR}{\overline{\abbrev{MDR}}\xspace}
\newcommand{\miss}{\text{miss}}
\newcommand{\MRSSM}{\abbrev{MRSSM}\xspace}
\newcommand{\MS}{\ensuremath{M_S}}
\newcommand{\MSbar}{\overline{\abbrev{MS}}\xspace}
\newcommand{\MSSM}{\abbrev{MSSM}\xspace}
\newcommand{\mueff}{\ensuremath{\mu_{\text{eff}}}}
\newcommand{\NLL}{\abbrev{NLL}}
\newcommand{\NMSSM}{\abbrev{NMSSM}\xspace}
\newcommand{\NMSSMCalc}{\texttt{NMSSMCalc}\xspace}
\newcommand{\NNLL}{\abbrev{N$^2$LL}}
\newcommand{\NNNLL}{\abbrev{N$^3$LL}}
\newcommand{\NLO}{\abbrev{NLO}}

\newcommand{\NNNLO}{\abbrev{N$^3$LO}}

\newcommand{\order}[1]{\ensuremath{\mathcal{O}\!\left(#1\right)}}
\newcommand{\QCD}{\abbrev{QCD}}

\newcommand{\Qin}{\ensuremath{Q_\abbrev{in}}}
\newcommand{\Qlow}{\ensuremath{Q_\abbrev{low}}}
\newcommand{\Qm}{\ensuremath{Q_{\abbrev{m}}}}
\newcommand{\Qpole}{\ensuremath{Q_\abbrev{pole}}}
\newcommand{\SA}{\texttt{SARAH}\@\xspace}
\newcommand{\SP}{\texttt{SPheno}\@\xspace}
\newcommand{\SLHA}{\abbrev{SLHA}\xspace}
\newcommand{\SM}{\abbrev{SM}\xspace}
\newcommand{\SUSY}{\abbrev{SUSY}\xspace}
\newcommand{\SusyHD}{\texttt{SusyHD}\xspace}

\newcommand{\tree}{{0\ell}}
\newcommand{\unit}[1]{\ensuremath{\,\text{#1}}}

\author[a]{Thomas Kwasnitza}
\author[a]{Dominik Stöckinger}
\author[b]{Alexander Voigt\orcidlink{0000-0001-8963-6512}}
\author[a]{Johannes Wünsche}
\affil[a]{Institut für Kern- und Teilchenphysik, TU Dresden, Zellescher Weg 19, 01069 Dresden, Germany}
\affil[b]{Institute for Theoretical Solid State Physics, RWTH Aachen University, Sommerfeldstra{\ss}e 16, 52074 Aachen, Germany}

\begin{document}
\maketitle
\begin{abstract}
We perform an extensive analysis of the light CP-even
Higgs boson pole mass in the \MSSM\ and its dependencies
on various parameters based on the 3-loop \FEFTH\ hybrid
calculation which is implemented and publicly available since
recently in \FS. Our focus lies on the study of the
robustness of the approach in scenarios of highly non-degenerate
\SUSY\ mass spectra. Also, we present an improved Higgs mass
calculation in the \NMSSM\ based on the same approach, which is
published in the new version 2.9.0 of \FS as well. The calculation provides a
treatment in the full-model parametrization, leading to an
advantageous resummation of \QCD-enhanced terms in the stop-mixing
parameter and includes important 2-loop contributions as well as
3-loop \QCD\ contributions in the \MSSM\ limit.  We assess the
reliability of this new calculation by applying it to several distinct
\NMSSM\ scenarios. In this context, special attention is devoted to
the estimation of \NMSSM-specific theory uncertainty.
\end{abstract}

\clearpage
\tableofcontents
\clearpage
\section{Introduction}

The discovery of a Higgs boson $h$ by the \abbrev{ATLAS} and
\abbrev{CMS} experiments \cite{Aad:2012tfa,Chatrchyan:2012xdj}
represents a milestone in our understanding of the physics of
elementary particles, because it confirms the existence of
the Higgs field generating the masses
of elementary particles within the Standard Model of particle physics
(\SM). However, the \SM\ cannot explain the mass $M_h$ of the Higgs
boson itself, because it is related to the quartic Higgs coupling
$\lambdaSM$, which is a free parameter of the \SM. For this reason and
because the \SM\ cannot provide an explanation of phenomena such as
dark matter or neutrino masses, models beyond the Standard Model
(\BSM) have been proposed. Among them is the class of supersymmetric
models which are able to predict the order of magnitude of the Higgs
boson mass $M_h$. However, these models also incorporate numerous
new particles, which have not been directly observed
yet. For this reason, if these new \BSM\ particles exist, they either have masses
$\MS$ of the order of the electroweak scale or below,
$\MS\lesssim M_Z$, but are difficult to produce, or their masses
are beyond the currently accessible energy scale of high-energy
physics experiments (beyond the electroweak scale),
$\MS\gg M_Z$. For both scenarios, there exist well-developed methods to
predict $M_h$ with reasonable precision (see e.g.\
Ref.~\cite{Slavich:2020zjv} for a review of available methods):
\begin{itemize}
\item In scenarios where $\MS\sim M_Z$, so-called fixed-order
  (\FO) calculations, where quantum corrections up to a fixed loop
  order are taken into account, lead to a reasonable
  precision. However, the prediction of $M_h$ with this method is
  imprecise if the loop contributions not taken into account are
  large, which is the case when $\MS\gg M_Z$.
\item In scenarios where $\MS\gg M_Z$, so-called effective field
  theory (\EFT) calculations, where large logarithmic higher-order
  contributions of $\order{\ln^n(M_Z^2/\MS^2)}$ are taken into
  account at all orders $n$, leading to a reasonable precision.
  However, the prediction of $M_h$ within an effective field theory
  becomes imprecise if $\MS\lesssim M_Z$, because the method neglects
  contributions of $\order{v^2/\MS^2}$, where $v\approx 246\unit{GeV}$
  is the vacuum expectation value of the \SM\ Higgs doublet.
\end{itemize}
Since the mass scale $\MS$ of the new \BSM\ particles is unknown, it
is a-priori not clear which of the two methods (\FO\ or \EFT) to use
for a specific \BSM\ parameter point to obtain a precise prediction of
$M_h$. Thus, a ``hybrid'' method combining the virtues of both the \FO
and \EFT\ calculative approaches is very advantageous. There are
currently two different hybrid \FO/\EFT\ methods realized in several
implementations:
\begin{itemize}
\item The \FH\ hybrid method
\cite{Hahn:2013ria,Bahl:2016brp,Staub:2017jnp,Bahl:2018jom,Bahl:2020tuq,Bahl:2020mjy}: The
first implementation of a hybrid calculation of the \MSSM\ Higgs mass was part
of the \FH\ version published in 2013. It combines the radiative corrections to
the Higgs mass matrix obtained from the \FO\ calculation with the numerically determined
corrections from the \EFT\ approach. Double-counting is circumvented by
additionally subtracting contributions which are covered by both approaches.
\item The \FEFTH\ method \cite{Athron:2016fuq,Athron:2017fvs,Kwasnitza:2020wli}:
The hybrid method of \FEFTH, being proposed in 2016, follows another
approach. Its essence consists in the usage of the Higgs pole mass matching in order
to determine the quartic Higgs coupling $\lambdaSM$ in the \SM\ at $\MS$.
It naturally allows the inclusion of terms suppressed by powers of $v^2/\MS^2$.
Then, $\lambdaSM$ is evolved down to the electroweak scale where the Higgs
pole mass is calculated, analogously to the pure \EFT\ approach. Similar hybrid
Higgs mass calculations based on this approach are implemented in \SA/\SP
\cite{Staub:2017jnp} and \NMSSMCalc\ \cite{Borschensky:2024utz}.
\end{itemize}
The \FEFTH\ method was first introduced in Ref.~\cite{Athron:2016fuq}
at the 1-loop level and was published in \FS~1.7.0
\cite{Athron:2014yba}. It was applied to the (real,
flavour-conserving) \MSSM, but also to the \NMSSM, the \MRSSM\ and the
E$_6$SSM, and achieved a precision of \NLO+\LL. The 1-loop \FEFTH\
method was refined in \FS~2.0.0 \cite{Athron:2017fvs}, to avoid the
inclusion of higher order large logarithmic contributions, resulting
in a precision of \NLO+\NLL. In Ref.~\cite{Kwasnitza:2020wli} the
\FEFTH\ method was applied to the \MSSM\ together with the inclusion of
loop contributions from the literature up to (including) 2-loop and 3-loop
level in the so-called ``gaugeless limit'' where the electroweak gauge
couplings $g$ and $g'$ are set to zero. The achieved precision was thus of \NNNLO+\NNNLL\ in the strong
coupling \cite{Kwasnitza:2021idg}. Furthermore, a full-model parametrization
was used instead of an \EFT-parametrization, leading to a very beneficial
effective resummation of \QCD-enhanced terms leading in the stop
mixing parameter. The resulting calculation can be regarded as
state-of-the-art, see Ref.~\cite{Slavich:2020zjv}. For discussions of
2-loop corrections taking into account contributions involving the
electroweak gauge couplings, see Ref.~\cite{Borowka:2018anu,Bagnaschi:2019esc,Meuser:2023vyg,Bahl:2023ead}.

In the present paper, the application of \FEFTH to the \MSSM at 3-loop
level is further pursued by widening the focus from ``standard'' scenarios
with degenerate \SUSY\ mass spectra to generic scenarios where all
mass parameters can be different. As one of our main purposes, we
scrutinize the accuracy and reliability
of the predictions also for such more general parameter choices in order to
corroborate the robustness of the implemented calculation. This includes a
check of the numerical stability of the implementation and the smallness of
the theory uncertainty, where both rely on the correct
cancellation of logarithmic terms in the matching relations. Also, various parameter
dependences are revisited.

Our second focus lies on the \NMSSM. In the context of Higgs mass calculation,
this model has been analyzed extensively. However, only very few studies employed a
hybrid calculative method for that purpose. The first hybrid calculation in the
\NMSSM was presented in Ref.~\cite{Athron:2016fuq} implementing the \FEFTH\-method
as a part of \FS. It included a 1-loop matching and achieved a precision of \NLO+\NLL\
after the refinement in \FS~2.0.0 \cite{Athron:2017fvs}, as mentioned above. Another
hybrid calculation in the \NMSSM based on the same algorighmic idea was presented in
Ref.~\cite{Borschensky:2024utz} and implemented in \NMSSMCalc \cite{Borschensky:2024utz}.
It realizes a 2-loop matching
with \MSSM-like 2-loop matching corrections of $\order{g_3^2 y_t^4 + y_t^6}$
as well as mixed strong-electroweak corrections from Ref.~\cite{Bagnaschi:2019esc},
however with terms suppressed by powers of $v^2/M_S^2$ only included up to 1-loop level.

As a second purpose of this publication, we  present a newly implemented
application of the \FEFTH\ method to the \NMSSM\ with a 3-loop
matching, i.e.\ including 2-loop matching corrections of
$\order{g_3^2(y_t^4+y_b^4) + (y_t^2+y_b^2)^3 + (y_t^2 + y_\tau^2)^3}$
and the dominant 3-loop corrections of $\order{y_t^4 g_3^4}$. The latter are provided
by the module \Himalaya ~\cite{Harlander:2017kuc,Harlander:2018yhj} which uses different
approximations from \cite{Kant:2010tf}, selected according to the considered hierarchy
between the gluino and the squark masses. This extension of the \FEFTH\ calculation in
the \NMSSM\ is publicly available since the recent version~2.9.0 of \FS and represents a
code achieving state-of-the-art precision for the Higgs mass calculation in the
\NMSSM. Similar to the \MSSM case, several scenarios and parameter dependences
are analysed, and the theory uncertainty of the calculation is
quantified by employing several 
methods of estimation.

\section{The FlexibleEFTHiggs method}

In this study we consider a calculation of the light CP-even Higgs
pole mass $M_h$ in the \MSSM and the \NMSSM. We employ a hybrid calculation
that resums large logarithmic contributions of the form
$\ln^n(M_Z^2/\MS^2)$ and at the same time includes complete series of
terms of $\order{v^n/\MS^n}$. For our calculation we use the refined
\FEFTH\ method described in detail in Ref.~\cite{Kwasnitza:2020wli}
and published in \FS~2.9.0.

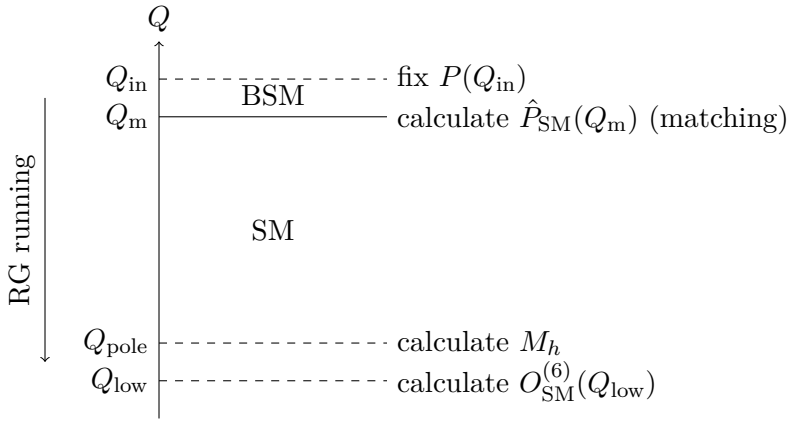
\begin{figure}
  \centering
  \begin{tikzpicture}
    \draw[->] (0,0) -- (0,5) node[above]{$Q$};
    \draw[dashed] (0,4.5) node[left]{$\Qin$} -- ++ (3,0) node[right]{fix $P(\Qin)$};
    \draw[] (0,4) node[left]{$\Qm$} -- node[above=0.5]{\BSM} ++ (3,0) node[right]{calculate $\hat{P}_\SM(\Qm)$ (matching)};
    \draw[dashed] (0,1) node[left]{$\Qpole$} -- ++ (3,0) node[right]{calculate $M_h$};
    \draw[dashed] (0,0.5) node[left]{$\Qlow$} -- ++ (3,0) node[right]{calculate $O_\SM^{(6)}(\Qlow)$};
    \draw[dashed] (1.5,2.5) node[]{\SM};
    \draw[<-] (-1.5,0.75) -- node[above,rotate=90]{RG running} ++ (0,3.5);
  \end{tikzpicture}
  \caption{Schematic illustration of the scales used in the refined
    \FEFTH\ method.}
  \label{fig:method}
\end{figure}
The \FEFTH\ method is schematically illustrated in
Figure~\ref{fig:method}. It is structurally similar to a pure \EFT\
calculation, where all \BSM\ particles are integrated out at a common
scale $\Qm$ of the order of the \BSM\ particle masses $\MS$, resulting
in the \SM\ as \EFT. The obtained \SM\ parameters are evolved down to
the electroweak scale $\Qlow\approx M_Z$, using $\SM$-renormalization group
running, where they are matched to known low-energy observables.  Due
to this running large logarithmic contributions to $M_h$ are
effectively resummed when the Higgs pole mass is calculated in the
\SM\ at the electroweak scale.
The special feature of the \FEFTH\ method lies in the matching
conditions between the \BSM\ model and the \SM: To determine the
quartic Higgs coupling $\lambdaSM(\Qm)$ in the \SM, a Higgs pole
mass matching condition of the form
\begin{equation}
  (M_h^\MSSM)^2 = (M_h^\SM)^2
\end{equation}
is used. When the squared \SM\ Higgs pole mass is written as
\begin{equation}
  (M_h^\SM)^2 = \lambdaSM(\Qm)\hat{v}^2 + \Delta s_h^{\SM},
\end{equation}
where $\Delta s_h^{\SM}$ are the \SM\ loop corrections, one obtains
the quartic Higgs coupling as
\begin{equation}
  \lambdaSM(\Qm) = \frac{1}{\hat{v}^2}\left[(M_h^\MSSM)^2 -  \Delta s_h^{\SM}\right].
  \label{eq:Mh-matching}
\end{equation}
In the \FEFTH\ method the r.h.s.\ of Eq.~\eqref{eq:Mh-matching} is
expanded in loops and couplings, but not in powers of $\hat{v}/\MS$.
In this way certain series of terms of the form $\order{v^n/\MS^n}$
are included in the value $\lambdaSM(\Qm)$. When the Higgs pole
mass is eventually calculated at the electroweak scale in the \SM\
using the value of $\lambdaSM(\Qpole\approx M_t)$, both large
logarithmic terms and terms of the form $\order{v^n/\MS^n}$ are
included.

In the following we describe the \FEFTH\ method to calculate $M_h$ in
more detail. While Ref.~\cite{Kwasnitza:2020wli} describes the method
for the \MSSM, we generalize it here to non-minimal supersymmetric
models, such as the \NMSSM. In the refined \FEFTH\ method the
prediction of the Higgs pole mass $M_h$ starts by specifying the
parameters $P(\Qin)$ of the full (supersymmetric) \BSM\ model in the
$\DRbarPrime$ scheme\footnote{The $\DRbarPrime$ scheme differs from naive
minimal substraction in the DR scheme by the renormalization of the
unphysical $\epsilon$-scalar mass. See \cite{Jack:1994rk} for details.}
at some user-chosen input renormalization scale
$\Qin$, see Figure~\ref{fig:method}. We divide these parameters into
the following two subsets, $P(\Qin)=P_{\SM}(\Qin)\cup P_{\BSM}(\Qin)$:
\begin{itemize}
\item \BSM\ parameters that have an \SM\ counterpart. These are the
  three electroweak gauge couplings $g_1$, $g_2$ and $g_3$, the third
  generation Yukawa couplings $y_t$, $y_b$ and $y_\tau$ and the
  \SM-like vacuum expectation value $v$.  We denote these parameters
  as
  \begin{equation}
    P_{\SM}(\Qin) = \left\{g_1(\Qin),g_2(\Qin),g_3(\Qin),y_t(\Qin),y_b(\Qin),y_\tau(\Qin),v(\Qin)\right\}.
    \label{eq:P_SM}
  \end{equation}
  These parameters are not free input parameters, because they will be
  fixed by measured low-energy quantities, see below.
\item Pure \BSM\ parameters $P_{\BSM}(\Qin)$, that have no \SM\
  counterpart. These are soft supersymmetry-breaking parameters,
  non-\SM\ superpotential parameters (such as the $\mu$ parameter in
  the \MSSM), non-\SM\ vacuum expectation values (such as $v_s$ in the
  \NMSSM) and combinations thereof (such as $\tan\beta=v_u/v_d$ in the
  \MSSM) etc. These parameters are free input parameters and are not
  fixed by measurements.
\end{itemize}
All parameters $P(\Qin)$ of the \BSM\ model are evolved to a
the matching scale $\Qm$ using the full 2-loop
renormalization group equations from
Refs.~\cite{Jones:1974pg,Jones:1983vk,West:1984dg,Martin:1993yx,Yamada:1993ga,Jack:1994kd,Jack:1994rk,Yamada:1994id,Martin:1993zk,Fonseca:2011vn,Goodsell:2012fm,Sperling:2013eva,Sperling:2013xqa}
with additional 3-loop \MSSM\ contributions from
Refs.~\cite{Jack:2003sx,Jack:2004ch}, if appropriate. The matching
scale $\Qm$ should be chosen to be of the order of the non-\SM\ \BSM\
particle masses $\MS$, i.e.\ $\Qm\sim\MS$.

At the single matching scale $\Qm$, all \BSM\ particles are integrated out
leaving the \SM\ as effective theory with a prediction of the \SM\ $\MSbar$ parameters
\begin{equation}
  \hat{P}_{\SM}(\Qm) = \left\{\hat{g}_1(\Qm),\hat{g}_2(\Qm),\hat{g}_3(\Qm),\hat{y}_t(\Qm),\hat{y}_b(\Qm),\hat{y}_\tau(\Qm),\hat{v}(\Qm),\lambdaSM(\Qm)\right\}
  \label{eq:Phat_SM}
\end{equation}
from the \BSM\ parameters $P(\Qm)$. The imposed matching conditions
are described in detail in Ref.~\cite{Kwasnitza:2020wli}. An important
point w.r.t.\ this matching is the employed Higgs pole mass matching
condition, which allows the inclusion of terms suppressed by powers of
$v^2/\MS^2$ in the quartic \SM\ Higgs coupling $\lambdaSM(\Qm)$
in a natural way. The matching conditions are solved perturbatively for the \SM\
$\MSbar$ parameters \eqref{eq:Phat_SM} at the matching scale
in terms of loops, as described in Ref.~\cite{Kwasnitza:2020wli}. For
instance, for the \SM\ quartic $\MSbar$ Higgs coupling $\lambdaSM$
one obtains
\begin{equation}
  \lambdaSM(\Qm) = \frac{m_h^2(\Qm)}{v^2(\Qm)} + \DeltalambdaSM^{1\ell}(\Qm) + \DeltalambdaSM^{2\ell}(\Qm) + \DeltalambdaSM^{3\ell}(\Qm),
  \label{eq:lambda-hat}
\end{equation}
where $m_h$ is the \SM-like $\DRbarPrime$ Higgs mass in the \BSM\
model and $\DeltalambdaSM^{k\ell}$ are the $k$-loop threshold
corrections. In all \BSM\ models the complete 1-loop threshold
corrections are taken into account, including regularization
conversion terms.  In the \MSSM\ and \NMSSM\ we include threshold
corrections of 2-loop order
$\order{g_3^2(y_t^4+y_b^4) + (y_t^2+y_b^2)^3 + (y_t^2 + y_\tau^2)^3}$
from
Refs.~\cite{Degrassi:2001yf,Brignole:2001jy,Dedes:2002dy,Brignole:2002bz,Dedes:2003km}
and of 3-loop order $\order{g_3^4y_t^4}$ from
Ref.~\cite{Harlander:2018yhj}.
There are two important features that distinguish
Eq.~\eqref{eq:lambda-hat} from pure effective field theory (\EFT)
calculations such as \SusyHD\ \cite{PardoVega:2015eno} or \HSSUSY\
\cite{Athron:2017fvs,Harlander:2018yhj}:
\begin{itemize}
\item The matching calculation, including e.g.\ the r.h.s.\
  of Eqs.~\eqref{eq:Mh-matching} or~\eqref{eq:lambda-hat} is
  expressed as a perturbative expansion in terms of the BSM model
  parameters, i.e.\ not in terms of the \SM\ parameters. In
  Ref.~\cite{Kwasnitza:2020wli} this approach was therefore called
  ``full-model parametrization''. It has been shown in
  Ref.~\cite{Kwasnitza:2021idg} that this full-model
  parametrization leads to an effective resummation of QCD-enhanced terms that
  involve (potentially large) trilinear squark--Higgs couplings in the
  \MSSM. The same resummation effect is also present in
  the \NMSSM. At the same time, the full-model parametrization substantially
  simplifies the application of the computational setup to non-minimal SUSY
  models such as the NMSSM or beyond.

\item In the derivation of Eq.~\eqref{eq:lambda-hat} no expansion in
  powers of $v^2/\MS^2$ is performed. Therefore $\lambdaSM(\Qm)$
  contains terms of $\order{v^2/\MS^2}$ in an unexpanded form, which
  are not included in the pure-\EFT\ calculations of \SusyHD\ and
  \HSSUSY. These terms are of importance in parameter scenarios that
  contain \BSM\ particles with masses not large compared to the
  electroweak scale, i.e.\ where $v\sim\MS$.  The inclusion of these
  terms leads to the specific feature that the Higgs boson pole mass
  calculated with the hybrid \FEFTH\ method converges to the fixed-order
  calculation for $\MS\to M_Z$ and to the pure-\EFT\
  calculation for $\MS\to\infty$ (as long as both calculations are
  expressed in terms of the same parameters).
\end{itemize}
The \SM\ $\MSbar$ parameters $\hat{P}_{\SM}(\Qm)$ from
Eq.~\eqref{eq:Phat_SM}, which have been determined from the
matching of the \BSM\ model to the \SM, are evolved to the electroweak
scale $\Qlow=M_Z$ using the \SM\ renormalization group equation with
contributions up to the 4-loop level from
Refs.~\cite{Mihaila:2012fm,Bednyakov:2012rb,Bednyakov:2012en,Chetyrkin:2012rz,Bednyakov:2013eba,Chetyrkin:2016ruf,Martin:2015eia,Bednyakov:2015ooa}. The
so-obtained \SM\ $\MSbar$ parameters $\hat{P}_{\SM}(\Qlow)$ are then
used to calculate the following \SM\ quantities:
\begin{equation}
  O_{\SM}^{(6)}(\Qlow) = \left\{\aem^{\SM(6)}(\Qlow),\as^{\SM(6)}(\Qlow),G_F,M_t,M_Z,m_b^{\SM(6)}(\Qlow),M_\tau\right\}.
  \label{eq:observables}
\end{equation}
In Eq.~\eqref{eq:observables} $\aem^{\SM(6)}(\Qlow)$ and
$\as^{\SM(6)}(\Qlow)$ denote the $\MSbar$ electromagnetic and the
strong couplings in the \SM\ with six active quark flavours,
respectively, $G_F$ is the Fermi constant, $M_t$ is the top quark pole
mass, $M_Z$ is the $Z$ boson pole mass, $m_b^{\SM(6)}(\Qlow)$ is the
bottom quark $\MSbar$ mass in the \SM\ with six active quark flavours
and $M_\tau$ is the $\tau$ lepton pole mass. The loop orders, at which
these quantities are calculated, are chosen such that the desired
precision of the prediction of $M_h$ is reached, see
Ref.~\cite{Kwasnitza:2020wli} for details.
The electromagnetic and strong couplings, as well as the $\MSbar$
bottom quark mass from Eq.~\eqref{eq:observables} are further
converted to their $\MSbar$ counterparts in the \SM\ with only five
active quark flavours. One thus arrives at the following set of
low-energy quantities:
\begin{equation}
  O_{\SM}^{(5)} = \left\{\aem^{\SM(5)}(M_Z),\as^{\SM(5)}(M_Z),G_F,M_t,M_Z,m_b^{\SM(5)}(m_b),M_\tau\right\}.
  \label{eq:observables-2}
\end{equation}
The so-obtained low-energy quantities from
Eq.~\eqref{eq:observables-2} are compared to corresponding user-given
low-energy input parameters.  If the two parameter sets are not equal,
the \SM-like \BSM\ parameters $P_{\SM}(\Qin)$ in Eq.~\eqref{eq:P_SM}
are varied until the two parameter sets match. This procedure is
denoted as ``shooting solver'' in \FS.

Once a parameter set $P_\SM(\Qin)$ has been determined that is
consistent with the user-given input the Higgs boson pole mass $M_h$
is calculated in the \SM\ in the $\MSbar$ scheme at the
renormalization scale $\Qpole=M_t$ using the parameters
$\hat{P}_\SM(\Qpole)$. In this calculation we include the full 1-loop
contributions, the 2-loop contributions of the order
$\order{\hat{g}_3^2(\hat{y}_t^4\hat{v}^2+\hat{y}_b^4) +
  (\hat{y}_t^2+\hat{y}_b^2)^3\hat{v}^2 + (\hat{y}_t^2 +
  \hat{y}_\tau^2)^3\hat{v}^2}$ from
Refs.~\cite{Degrassi:2012ry,Martin:2014cxa,PardoVega:2015eno,Kwasnitza:2020wli}
and the 3-loop contributions of
$\order{\hat{g}_3^4\hat{y}_t^4\hat{v}^2}$ from
Ref.~\cite{Martin:2014cxa}.
The precision of this Higgs pole mass prediction is thus of \NNNLO\
with a large-log resummation\footnote{Due to the single matching scale,
at which all \BSM\ particles are integrated out, the resummation affects
only large logarithms $\ln(M_Z/\MS)$ while logarithms of \BSM\ mass
ratios are not resummed but only captured at the considered fixed loop
order of the high-scale matching calculation.}
of \NNNLL\ in the product of couplings
given above. In the \MSSM\ this precision has been reached before in
Refs.~\cite{Kwasnitza:2020wli,Harlander:2019dge}. In the \NMSSM,
however, this precision has not been reached before, to the best of
our knowledge.

\section{Application to the \MSSM}

The refined \FEFTH\ method with the precision at \NNNLO\ and \NNNLL\
resummation as well as $x_t$-resummation
has been described for the CP-conserving \MSSM\ in detail
in Ref.~\cite{Kwasnitza:2020wli}.  However, in this reference the
application of the method to concrete scenarios was restricted to
simplified ``standard'' scenarios with degenerate \SUSY\ mass
parameters. In this section, we extend the application to various
non-trivial scenarios with non-degenerate \SUSY\ mass parameters.

\subsection{Model definition}

We very briefly introduce some conventions used for the following
discussion of the CP-conserving \MSSM. The utilized conventions are
those used in Ref.~\cite{Kwasnitza:2020wli}. We define all \MSSM\
parameters $P(Q)=P_{\SM}(Q)\cup P_{\MSSM}(Q)$ in the $\DRbarPrime$
scheme.  The \SM-like \MSSM\ parameters $P_{\SM}(Q)$ are denoted as in
Eq.~\eqref{eq:P_SM}. By \EWSB, the neutral components of the two
\MSSM\ Higgs doublets $h_u$ and $h_d$ acquire vacuum expectation
values (VEVs), which we denote as $v_u$ and $v_d$, defined as
\begin{equation}
  \langle h_u \rangle = \frac{1}{\sqrt{2}} \begin{pmatrix} 0 \\ v_u \end{pmatrix}
  \qquad \text{and} \qquad
  \langle h_d \rangle = \frac{1}{\sqrt{2}} \begin{pmatrix} v_d \\ 0 \end{pmatrix}.
  \label{eq:Higgs_double_VEVs}
\end{equation}
Accordingly, $v_u$ and $v_d$ represent the true minimum of the
effective Higgs potential. Their ratio defines the tangent of the
angle $\beta$ via $\tan \beta = v_u/v_d$. Among the electrically
neutral physical Higgs states is one CP-odd Higgs boson $A$ and two
CP-even Higgs bosons $h$ and $H$ with $m_h\leq m_H$. The $\DRbarPrime$
mass of the lighter state $h$ is given by
\begin{align}
  m_h^2
  &= \frac{1}{2}\left[ m_Z^2 + m_A^2 - \sqrt{\left(m_Z^2 + m_A^2\right)^2 - 4 m_Z^2 m_A^2 \cos^2(2\beta)} \right] \\
  &= m_Z^2 \cos^2(2\beta) + \order{\frac{m_Z^4}{m_A^4}} \\
  &= \frac{1}{4} \left( g_Y^2 + g_2^2 \right) v^2 \cos^2(2\beta) + \order{\frac{v^4}{m_A^4}}
\end{align}
with $g_Y = \sqrt{5/3}\,g_1$ and
$v^2 = v_u^2 + v_d^2\approx (246\unit{GeV})^2$. The loop
contributions between $m_h$ and the corresponding pole mass $M_h$ are
sensitive to the mixing among the third generation sfermions.  This
mixing is governed by the mixing parameters $X_f$, $f\in\{t,b,\tau\}$,
which we define as
\begin{equation}
  X_t = A_t - \mu \cot\beta, \qquad
  X_b = A_b - \mu \tan\beta, \qquad
  X_{\tau} = A_{\tau} - \mu \tan\beta
  \label{eq:Xf}
\end{equation}
for the third generation sfermions $\tilde{f}_i$, $i\in\{1,2\}$. Here,
$\mu$ is the superpotential parameter describing the coupling between
the Higgs doublet superfields $\hat{H}_u$ and $\hat{H}_d$. The $A_f$
are the trilinear Higgs--sfermion--sfermion couplings. In addition, we
define the dimensionless third generation sfermion mixing parameters
\begin{equation}
x_t = \frac{X_t}{\left[(m_{\tilde{q}}^2)_{33}(m_{\tilde{u}}^2)_{33}\right]^{1/4}}, \quad
x_b = \frac{X_b}{\left[(m_{\tilde{q}}^2)_{33}(m_{\tilde{d}}^2)_{33}\right]^{1/4}}, \quad
x_{\tau} = \frac{X_{\tau}}{\left[(m_{\tilde{l}}^2)_{33}(m_{\tilde{e}}^2)_{33}\right]^{1/4}},
\label{eq:xt}
\end{equation}
where $m_{\tilde{q}}^2$, $m_{\tilde{u}}^2$ and $m_{\tilde{d}}^2$ are
the soft-breaking squared mass parameter matrices of the left- and
right-handed quark superpartners, respectively. Likewise, the
soft-breaking squared mass parameter matrices for the left- and
right-handed lepton superpartners are represented by $m_{\tilde{l}}^2$ and
$m_{\tilde{e}}^2$, respectively. We denote the soft-breaking gaugino
mass parameters as $M_i$, $i\in\{1,2,3\}$, where $M_1$ corresponds to
the bino, $M_2$ corresponds to the wino and $M_3$ corresponds to the
gluino mass. We use the symbol $\mu$ for \MSSM's superpotential
$\mu$-parameter. Finally, we use the symbol $\MS$ to denote a \SUSY\
mass scale parameter, which describes the order of the non-SM
BSM particle masses.

\subsection{Uncertainty estimation}
\label{sec:MSSM-UncEst}

In order to assess the reliability of the Higgs pole mass prediction,
we estimate its uncertainty in a similar manner as in
Ref.~\cite{Kwasnitza:2020wli}. Accordingly, we consider two
contributions to the total uncertainty: a ``high-scale uncertainty''
to estimate missing contributions in the matching of the \MSSM\ to the
\SM, and a ``low-scale uncertainty'' to estimate missing contributions
in the calculation of the Higgs pole mass in the \SM\ at the scale
$\Qpole$ and in the calculation of the low-scale quantities
$O_\SM^{(5)}$. Due the imposed pole mass matching conditions, series
of power-suppressed terms of $\order{v^2/\MS^2}$ are included in the
calculation by construction at the full 1-loop and Yukawa- and
\QCD-enhanced 2-loop level. We do not include a so-called ``EFT
uncertainty'' that would correspond to missing power-suppressed terms
(for example power-suppressed 2-loop electroweak terms or non-resummed
power-suppressed large logarithms from the lack of running dimension-6
operators). The impact of such missing terms is expected to be
significantly smaller than the high-scale uncertainty from e.g.
non-power-suppressed missing electroweak corrections.

\subsubsection{High-scale uncertainty}

As in Ref.~\cite{Kwasnitza:2020wli}, we estimate the high-scale
uncertainty by varying the matching scale $\Qm\in[\Qin/2,2\Qin]$
while keeping $\Qin$ unchanged. We take the resulting variation
of $M_h$ as uncertainty estimate,
\begin{equation}
  \Delta M_h^{\Qm} = \max_{Q \in [\Qin/2,2\Qin]} \Big\{\big| M_h(\Qm=\Qin) - M_h(\Qm=Q) \big|\Big\}.
  \label{eq:DMh_Qm}
\end{equation}
This method is commonly used in the literature
\cite{Bagnaschi:2014rsa,PardoVega:2015eno,Allanach:2018fif,Bahl:2019hmm,Bahl:2019wzx,Borschensky:2024utz}
and well-proven to be a reliable approach to estimate the
missing \MSSM-like higher order terms, at least for moderately large
values of the stop mixing parameter ($|x_t|\leq3$), see
Ref.~\cite{Kwasnitza:2020wli}. Here, we desist from additionally
estimating the high-scale uncertainty by regarding implicit
corrections at higher order and reparametrization terms, as this was
necessary in Ref.~\cite{Kwasnitza:2020wli} to cover missing terms of
highest order in $x_t$. For $|x_t|\leq3$, the high-scale
uncertainty estimation via matching scale variation turned out to be
sufficient. Since only moderate values of $x_t$ are considered here
and, furthermore, the study of the $x_t$-dependence is not the focus
of this work, the procedure can accordingly be simplified.

\subsubsection{Low-scale uncertainty}

In accordance with Ref.~\cite{Kwasnitza:2020wli}, we employ two
methods to estimate the low-scale uncertainty: The variation of the
renormalization scale $\Qpole$ for the calculation of the Higgs pole
mass $M_h$ in the \SM\ and the variation of the loop order at which
the top quark pole mass $M_t$ is calculated in the \SM\ from the
$\MSbar$ parameters $\hat{P}_{\SM}(\Qlow)$.

\paragraph{Pole mass scale variation.} In analogy to the matching
scale variation, we vary the pole mass scale $\Qpole$, at which the
Higgs pole mass $M_h$ is calculated in the \SM, between $M_t/2$ and
$2M_t$.  We take the resulting variation of the Higgs pole mass as
uncertainty estimate,
\begin{equation}
  \Delta M_h^{\Qpole} = \max_{Q \in [M_t/2,2M_t]} \left\{\left| M_h(\Qpole=M_t) - M_h(\Qpole=Q) \right|\right\}.
\end{equation}
The so-defined uncertainty $\Delta M_h^{\Qpole}$ is an estimate of
logarithmic renormalization scale dependent \SM\ contributions, not
included in our calculation of $M_h$.

\paragraph{Variation of top quark pole mass loop order.} Another
source of low-scale uncertainty is the truncation of the perturbation
series for the calculation of the low-scale quantities $O_\SM^{(5)}$.
These low-scale quantities implicitly fix the \SM\ parameters
$\hat{P}_{\SM}(\Qlow)\setminus\{\lambdaSM(\Qlow)\}$.  The largest
uncertainty can be attributed to the fixation of the \SM\ $\MSbar$ top
quark Yukawa coupling $\hat{y}_t$ from the top quark pole mass
$M_t$. For the desired Higgs pole mass prediction at \NNNLO\ with
\NNNLL\ resummation, 3-loop \QCD\ corrections must be included in the
relation between $M_t$ and $\hat{y}_t$. We denote the Higgs mass,
where up to 3-loop \QCD\ contributions to $M_t$ are included as
$M_h^{y_t,3\ell}$.  We then estimate the missing higher order
contributions by including the available 4-loop \QCD\ contribution
from Ref.~\cite{Martin:2016xsp} in the calculation of $M_t$ in the
\SM\ at the scale $\Qlow$. The resulting Higgs pole mass is denoted as
$M_h^{y_t,4\ell}$.  We then take the variation of the predicted Higgs
pole masses as uncertainty estimate
\begin{equation}
  \Delta M_h^{y_t} = \left| M_h^{y_t,3\ell} - M_h^{y_t,4\ell} \right|,
\end{equation}
which is in accordance with Ref.~\cite{Kwasnitza:2020wli}.

\subsubsection{Combined uncertainty}

In order to obtain an estimate of the high-scale uncertainty $\Delta M_h^{\HS}$, just one
method was employed, namely the matching scale variation. For the
low-scale uncertainty, two different approaches are applied. Since
there is a significant overlap between the estimated missing
higher-order terms in both methods, we proceed as in
Ref.~\cite{Kwasnitza:2020wli} and take the maximum of both estimates
as combined low-scale uncertainty $\Delta M_h^{\LS}$.
We eventually take the linear addition of the obtained high-scale and
low-scale uncertainties as a conservative estimation for the total
uncertainty $\Delta M_h$:
\begin{equation}
  \Delta M_h = \Delta M_h^{\LS} + \Delta M_h^{\HS} = \max \left\{ \Delta M_h^{\Qpole}, \Delta M_h^{y_t} \right\} + \Delta M_h^{\Qm}.
  \label{eq:Comb_Uncert}
\end{equation}
This approach is again in accordance with
Ref.~\cite{Kwasnitza:2020wli}.

\subsection{Parameter dependencies}

In the past years, extensive scans have been performed to investigate
the Higgs mass dependence on the numerous parameters of the
\MSSM. Mainly, the regarded scenarios incorporated ``standard'' spectra
with degenerate or nearly degenerate soft-breaking mass parameters,
like the ones presented in Ref.~\cite{Slavich:2020zjv}. Such standard
scenarios have also been studied with the 3-loop
\FEFTH\ calculation in Ref.~\cite{Kwasnitza:2020wli}. In the
following we extend the consideration by including scenarios with
non-standard spectra. These scans also serve as a further practical
test of the stability of the \FEFTH\ calulation.

If not stated otherwise, we set the following \MSSM\ $\DRbarPrime$
parameters to a common \SUSY\ mass scale $\MS$ parameter in our
analysis:
\begin{subequations}
  \begin{align}
    (m_{\tilde{f}}^2)_{ij}(\Qin) &= \MS^2\,\delta_{ij}, & & f\in\{q,u,d,l,e\}, \ i\in\{1,2,3\}, \label{eq:parameter_condition_MSSM_mf} \\
    M_i(\Qin) &= \MS, & & i\in\{1,2,3\}, \label{eq:parameter_condition_MSSM_Mi} \\
    A_f(\Qin) &= 0, & & f\in\{u,d,c,s,b,e,\mu,\tau\}, \label{eq:parameter_condition_MSSM_Af} \\
    \mu(\Qin) = m_A(\Qin) &= \MS.
  \end{align}\label{eq:parameter_condition_MSSM}%
\end{subequations}
We set the input scale $\Qin$ equal to the matching scale $\Qm$ and
equal to the \SUSY\ scale parameter $\MS$, i.e.\ $\Qin=\Qm=\MS$. In
our study, we vary the stop mixing parameter $x_t(\Qin)$ and fix
$A_t(\Qin)$ according to Eqs.~\eqref{eq:Xf}--\eqref{eq:xt}. The
low-scale parameters $O_{\SM}^{(5)}$ are set to
\begin{subequations}
  \begin{align}
    \aem^{\SM(5)}(M_Z) &= 1/127.916, \\
    \as^{\SM(5)}(M_Z) &= 0.1184, \\
    G_F &= 1.1663787\cdot 10^{-5}\unit{GeV}^{-2}, \\
    M_t &= 173.34\unit{GeV}, \\
    M_Z &= 91.1876\unit{GeV}, \\
    m_b^{\SM(5)}(m_b) &= 4.18\unit{GeV}, \\
    M_\tau &= 1.777\unit{GeV}.
  \end{align}
\end{subequations}

\subsubsection{Dependence on \boldmath$M_3$}

\begin{figure}
  \begin{subfigure}[b]{0.5\textwidth}
    \includegraphics[width=\textwidth]{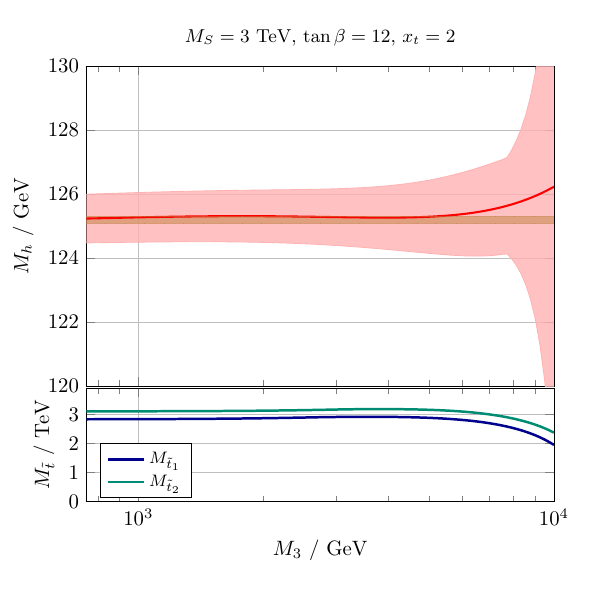}
  \end{subfigure}
  \begin{subfigure}[b]{0.5\textwidth}
    \includegraphics[width=\textwidth]{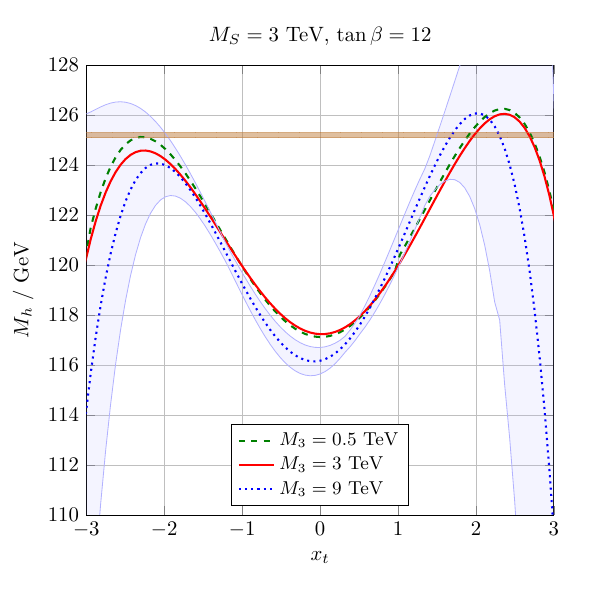}
  \end{subfigure}
  \caption{Left panel: Dependence of $M_h$ (and the stop pole masses
    $M_{\tilde{t}_{1,2}}$) on the gluino mass
    parameter $M_3$ with all other parameters fixed as in
    Eqs.~\eqref{eq:parameter_condition_MSSM}. The kink in the red
    uncertainty band at $M_3 \approx 8\unit{TeV}$ originates
    from the high-scale uncertainty determined according to
    Eq.~\eqref{eq:DMh_Qm}. At that point, the maximal difference is no
    longer found at the upper end of the $\Qm$ variation interval
    but at the lower one.
    Right panel: Dependence of $M_h$ on the stop mixing parameter
    $x_t$ for different values of $M_3$. For $M_3 = 9\unit{TeV}$,
    the uncertainty is drawn as light blue band.}
  \label{fig:Mh_M3}
\end{figure}

First we study the impact of the Higgs pole mass $M_h$ on the gluino
mass parameter $M_3$. In Figure~\ref{fig:Mh_M3} (left panel) we show
$M_h$ as a function of $M_3$ for the fixed values $\MS=3\unit{TeV}$,
$\tan\beta=12$ and $x_t=2$ (red solid line with red uncertainty
band). The brown band marks the experimental value
$M_h=(125.20 \pm 0.11)\unit{GeV}$ \cite{ParticleDataGroup:2024cfk}.
In the range $M_3\in[10^2,5\cdot 10^3]\unit{GeV}$ the
Higgs mass $M_h$ is nearly independent of $M_3$, up to deviations of
less than $1\unit{GeV}$. At the same time, the estimated uncertainty
$\Delta M_h$ remains essentially constant and of the order of
$\Delta M_h\approx 1\unit{GeV}$, underlining the robustness of the
used \FEFTH\ method in this range. However, for $M_3$ notably larger than
$\MS$, the calculated Higgs pole mass and especially its
uncertainty increase significantly. The rise in the Higgs mass is a
consequence of 2-loop corrections to $\lambdaSM(\Qm)$ of
$\order{g_3^2y_t^4}$, which contain terms quadratic in $M_3$ that stem
from gluino-induced stop self-energy corrections. A detailed analysis
of the terms can be found in Ref.~\cite{Degrassi:2001yf}.

The observed increased size of the uncertainty for large values of $M_3$ can be ascribed to
the high-scale uncertainty determined via matching
scale variation. As explained in Ref.~\cite{Bahl:2019hmm}, the
gluino mass parameter appears linearly in the 1-loop renormalization
group equation of $x_t$ and quadratically in those of
the squark mass parameters. Larger $M_3$ therefore imply
an enhancement of the respective \MSSM\ parameters during
renormalization group running. As a consequence, 
changes in $\Qm$ have a stronger impact on the numerical values
of these parameters. Due to the strong sensitivity of the Higgs
pole mass to $x_t$ and to the squark mass parameters, $M_h$ undergoes
large variations while varying $\Qm$, which results in a large
theory uncertainty. The large differences between the stop pole
masses and their corresponding squark mass parameters,
$(m_{\tilde{q}})_{33}$ and $(m_{\tilde{u}})_{33}$, visible for
$M_3 \gg \MS$ in Figure~\ref{fig:Mh_M3} (left panel)
indicate that a more physical renormalization scheme may be
very beneficial to stabilize the calculation. In Ref.~\cite{Kant:2010tf},
the so-called $\MDR$ renormalization scheme was suggested as an
appropriate scheme to alleviate the problem. In Ref.~\cite{Bahl:2019wzx},
this scheme was applied to an \EFT\ calculation of $M_h$ leading
to a substantial reduction of the estimated uncertainty.
Currently, a treatment of that kind is not implemented in {\FS}.

In Figure~\ref{fig:Mh_M3} (right panel), we show the dependence of $M_h$ as a
function of $x_t$ for different values of the gluino mass parameter
and fixed $\MS=3\unit{TeV}$ and $\tan\beta=12$. It shows that changes
in $M_3$ lead to a clear deformation of the curve of $M_h$ as a
function of $x_t$, including a
change in the extent of the asymmetry and a compression in
$x_t$-direction. Conversely, this implies a notable $M_3$-dependence
of the values of $x_t$
required to reproduce the experimental $M_h$. This interplay between
$M_3$, $x_t$ and $M_h$ can be traced
back to the 2-loop \QCD\ contributions to $M_h^2$, which are linear in $M_3$
and odd in $x_t$, thus being responsible for a stronger asymmetry with growing
$M_3$, c.f.\ Refs.~\cite{Degrassi:2001yf,Bagnaschi:2014rsa}. Also, the strong
$x_t$-dependence of the uncertainty for $M_3=9\unit{TeV}$ is remarkable,
especially the strong increase for $|x_t|>1.5$.
This again demonstrates that the effect of huge uncertainties for
large $M_3$ is closely connected to the value of $x_t$.

\subsubsection{Dependence on \boldmath$\mu$ and \boldmath$m_A$}
\label{Sec:MA}

\begin{figure}
  \begin{subfigure}[b]{0.5\textwidth}
    \includegraphics[width=\textwidth]{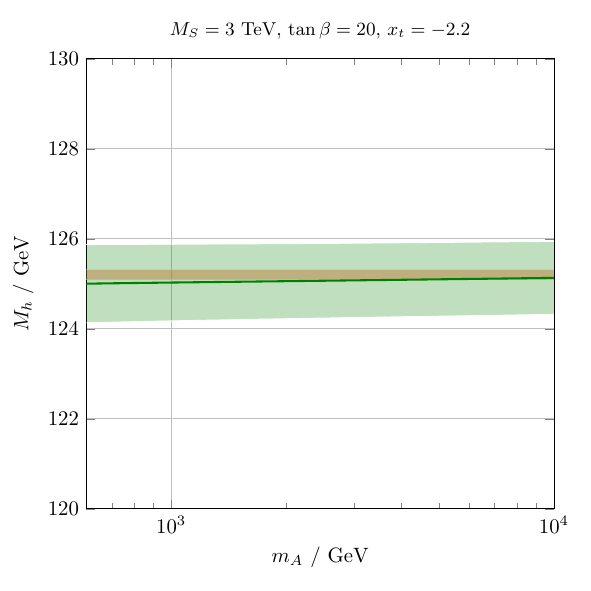}
  \end{subfigure}
  \begin{subfigure}[b]{0.5\textwidth}
    \includegraphics[width=\textwidth]{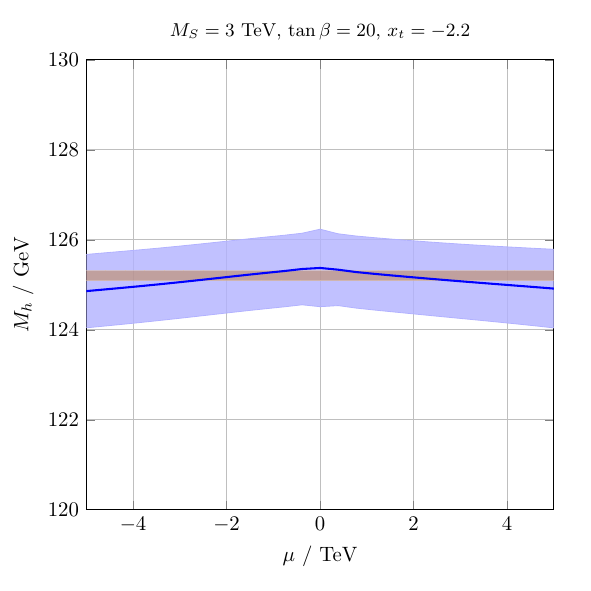}
  \end{subfigure}
  \caption{Left panel: $M_h$ as a function of the $\DRbarPrime$ CP-odd
    Higgs mass $m_A$. Right panel: $M_h$ as a function of the
    $\mu$-parameter. The parameters not varied are fixed to
    $\MS=3\unit{TeV}$, $\tan\beta=20$ and $x_t=-2.2$. The green band
    (left panel) and the blue band (right panel) show the uncertainty
    $\Delta M_h$. The brown band corresponds to the measured Higgs
    mass with its experimental uncertainty.}
  \label{fig:Mh_Mu_MA}
\end{figure}

In Figure~\ref{fig:Mh_Mu_MA}, we show the dependence of $M_h$ on the
$\DRbarPrime$ CP-odd \MSSM\ Higgs mass $m_A$ (left panel) and on the
$\mu$-parameter (right panel) for $\MS=3\unit{TeV}$, $\tan\beta=20$
and $x_t=-2.2$. We find that a variation of $m_A$, even by orders of
magnitude, has no large impact on $M_h$ for the chosen value of $\tan\beta$.
For smaller values of $\tan\beta$, a more relevant dependence can occur,
this would however also require significantly larger $\MS$ in order to
obtain sufficiently large corrections to $M_h$. In contrast, a
variation of $\mu$ between $-5\unit{TeV}$ and $5\unit{TeV}$ leads
to a mild change in $M_h$ of about $1\unit{GeV}$, where larger $|\mu|$
imply smaller $M_h$. These results are fully in accordance with
numerous analyses performed in the past
\cite{Chankowski:1992er,Dabelstein:1994hb,Carena:1995bx,Heinemeyer:1998np}.
For both dependencies, the corresponding uncertainty
remains nearly constant with a numerical value of about $0.8\unit{GeV}$.
Merely for large $\mu$, a slightly increased uncertainty can be
noticed.

We note that the shown parameter regions in Figure~\ref{fig:Mh_Mu_MA}
contain points with very split \SUSY\ mass spectra. At the same time
we find that the predictions of $M_h$ with the \FEFTH\ method is
numerically stable in these regions. Nontheless, in regions with
large, positive $x_t$ and $m_A \ll M_S$, negative $m_A^2$ may occur
during the renormalization group running and the \FEFTH\ calculation
breaks down. This issue has been discussed in
Ref.~\cite{Harlander:2019dge} in the context of a fixed-order
$\DRbarPrime$ calculation. To study such regions with tachyonic $m_A$
an effective field theory with a light $m_A$ could be beneficial.

\subsubsection{Dependence on \boldmath$(m_{\tilde{q}})_{33}$ and \boldmath$(m_{\tilde{u}})_{33}$}

\begin{figure}
 \begin{subfigure}[b]{0.5\textwidth}
 \includegraphics[width=\textwidth]{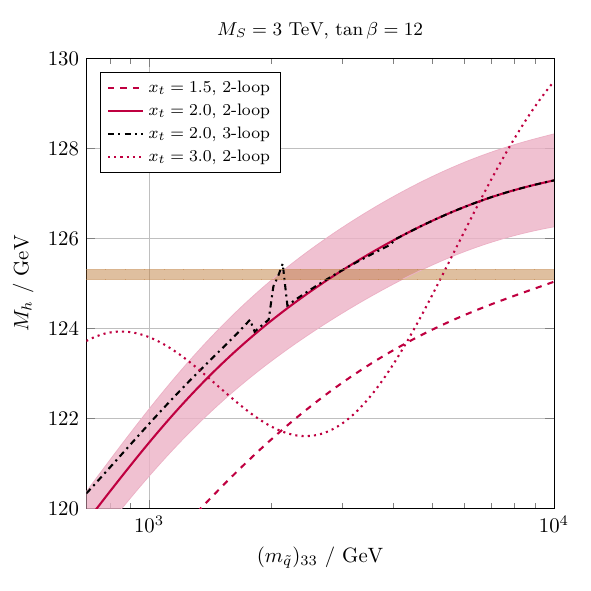}
 \end{subfigure}
 \begin{subfigure}[b]{0.5\textwidth}
 \includegraphics[width=\textwidth]{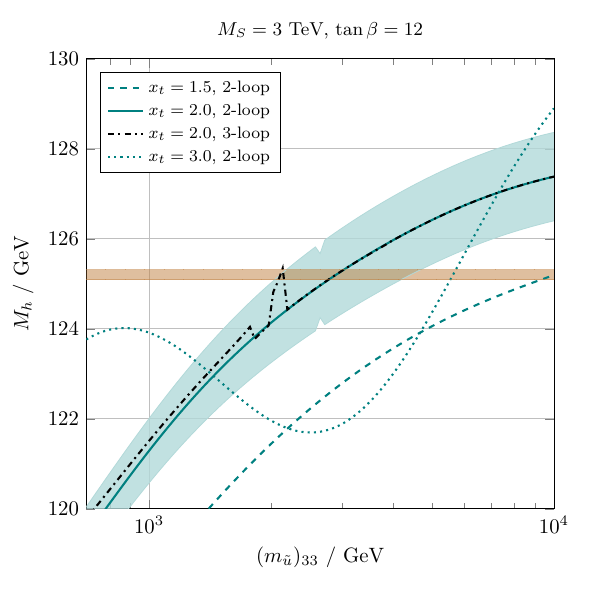}
 \end{subfigure}
 \caption{Left panel: $M_h$ as a function of $(m_{\tilde{q}})_{33}$ for
   different values of $x_t$. Right panel: $M_h$ as a function of
   $(m_{\tilde{u}})_{33}$ for different values of $x_t$. The parameters
   not varied are fixed to $\MS=3\unit{TeV}$ and $\tan \beta =
   12$. All curves are obtained from a 2-loop matching except
   the black dashed ones which represent the 3-loop results only for
   $x_t=2$. Also for $x_t=2$, the uncertainty bands (corresponding
   to the 2-loop calculation) are drawn in red (left panel) and teal
   (right panel), respectively. The brown band corresponds to the
   measured Higgs mass with its experimental uncertainty.}
\label{fig:Mh_mq33_mu33}
\end{figure}

In Figure~\ref{fig:Mh_mq33_mu33} we show the dependence of $M_h$ on
the sfermion mass parameters $(m_{\tilde{q}})_{33}$ (left panel) and
$(m_{\tilde{u}})_{33}$ (right panel) for different values of
$x_t$. Here, the strong impact of the induced change in the
$\DRbarPrime$ stop masses $m_{\tilde{t}_{1,2}}$ on the Higgs mass
$M_h$ becomes evident, which is much more significant than a change of
$M_3$, $m_A$ or $\mu$.
This high sensitivity originates from stop-induced 1-loop corrections to $M_h$.
The two plots reveal a strong similarity in the
dependence of $M_h$ on the two sfermion mass parameters. Especially,
the corresponding curves for the same fixed value of $x_t$ have a very similar shape.
Furthermore we find that $M_h$ can be maximized if $(m_{\tilde{q}})_{33}$
and $(m_{\tilde{u}})_{33}$ are chosen appropriately, again for a given
value of $x_t$.

It is important to note that the results given in the plots of
Figure~\ref{fig:Mh_mq33_mu33} represent the outcome of the \FEFTH
calculations with a matching at 2-loop level, except for the
black dashed lines, which were obtained from a 3-loop matching for
$x_t=2$. The shown uncertainty band is depicted for the 2-loop calculations for $x_t=2$.
The uncertainty is found to remain nearly constant within the
shown parameter range with numerical values slightly below
$1\unit{GeV}$. While the 2-loop curves and their uncertainties attest a
good stability of the \FEFTH\ calculation, the 3-loop results are subject to
several noteworthy jumps. This behaviour can be traced back to the 3-loop
threshold corrections for $\lambdaSM$ provided by \Himalaya
~\cite{Harlander:2017kuc,Harlander:2018yhj}, for which different approximations
are applied depending on the present hierarchy between gluino and squark masses.
Hierarchy transitions taking place during the variation of an individual
squark mass parameter then induce discontinuities in the numerical value
of $M_h$. Also, it must be emphasized that for
$(m_{\tilde{q}})_{33} \gtrsim 4\unit{TeV}$ there is no suitable
hierarchy implemented, leading to an absence of all 3-loop
contributions in the matching of the \BSM\ to the \SM. This statement
analogously holds for $(m_{\tilde{u}})_{33}$. However, the
discontinuity at that threshold is hardly visible in the plots.

This analysis indicates that the 3-loop calculation is stable and precise
in case of scenarios with degenerate mass spectra, especially for
$(m_{\tilde{q}})_{33} \sim M_S$ and $(m_{\tilde{u}})_{33} \sim M_S$.
However, it also reveals that in scenarios with non-degenerate squark
masses the calculation becomes unstable due to the mentioned hierarchy
transitions. In such cases, the 2-loop result can be regarded as more trustworthy.

\subsection{Benchmark scenarios}

Finally, we want to identify phenomenologically interesting regions of the
\MSSM parameter space which do not correspond to ``standard''
scenarios with degenerate mass parameters. For that porpose, a random scan is performed for which the \MSSM\ parameters
$P_{\BSM}(\Qin)$ are varied in the following ranges:
\begin{subequations}
  \begin{align}
    \MS &= (m_{\tilde{f}})_{11} = (m_{\tilde{f}})_{22} = 3\unit{TeV}, & 1 & \leq \tan\beta \leq 20, \\
    1\unit{TeV} &\leq (m_{\tilde{f}})_{33} \leq 5\unit{TeV}, & 1\unit{TeV} & \leq M_1,M_2,M_3 \leq 5\unit{TeV}, \\
    1\unit{TeV} &\leq \mu \leq 5\unit{TeV}, & 1\unit{TeV} & \leq m_A \leq 5\unit{TeV}, \\
    -10.5\unit{TeV} & \leq X_t \leq 10.5\unit{TeV}, & -135\unit{TeV} & \leq X_b \leq 135\unit{TeV}, \\
    A_f &= 0, & 
  \end{align}\label{eq:MSSM_parameter_scan}%
\end{subequations}
where
$\tilde{f}\in\{\tilde{q},\tilde{u},\tilde{d},\tilde{l},\tilde{e}\}$ and
$f\in\{u,d,c,s,e,\mu,\tau\}$. The soft-breaking trilinear Higgs coupling
parameters $A_t$ and $A_b$ are chosen according to the values of $X_t$
and $X_b$ using Eqs.~\eqref{eq:Xf}. Below, we will indicate the squark
mixing in each scenario with the dimensionless mixing parameters $x_t$
and $x_b$.

From all the generated parameter points, we first extracted all those which ensure
vacuum stability and reproduce approximately the experimental value of the light
\SM-like Higgs boson mass $M_h\approx 125\unit{GeV}$. It is well-known that $\tan \beta$ and
the three stop-sector parameters ($(m_{\tilde{q}})_{33}$, $(m_{\tilde{u}})_{33}$ and $x_t$)
strongly affect the Higgs mass $M_h$. Large $\tan \beta$, $(m_{\tilde{q}})_{33}$ and
$(m_{\tilde{u}})_{33}$ as well as $|x_t| \approx \sqrt{6}$ tend to increase $M_h$ significantly,
such that this pattern is also common among the parameter points reproducing the experimental
value. In detail, however,  the required
values of $\tan \beta$ and the three stop-sector parameters  depend on
the non-degeneracy within the stop sector and on all remaining
parameters.

One finding worth mentioning is that all experimentally viable parameter scenarios were found to have
$\tan \beta \gtrsim 6$, given $1\unit{TeV} \leq (m_{\tilde{f}})_{33} \leq 5\unit{TeV}$.
Furthermore, from our scans we found that in all these scenarios the stop mixing parameter $x_t$ lies in the range $1.3 \lesssim |x_t| \lesssim 3.2$. One should note that this behaviour with respect to
the parameters $\tan\beta$ and $x_t$ depends significantly on the chosen variation range
of $(m_{\tilde{f}})_{33}$.

\begin{table}
  \centering
  \begin{tabular}{ccccccccc}
    \toprule
    BMP                        & M0       & M1       & M2        & M3       & M4       & M5     \\ \midrule
    $\tan \beta$               & $12$     & $6$      & $20$      & $20$     & $8$      & $9$     \\
    $M_1$/GeV                  & $3000$   & $3000$   & $3000$    & $3000$   & $3000$   & $5000$  \\
    $M_2$/GeV                  & $3000$   & $1200$   & $1500$    & $3000$   & $3000$   & $4200$  \\
    $M_3$/GeV                  & $3000$   & $3000$   & $4000$    & $3000$   & $3000$   & $3000$  \\
    $m_A$/GeV                  & $3000$   & $3000$   & $3000$    & $3000$   & $3000$   & $4800$  \\
    $\mu$/GeV                  & $3000$   & $1500$   & $1200$    & $3000$   & $3000$   & $3000$  \\
    $(m_{\tilde{q}})_{33}$/GeV & $3000$   & $4500$   & $4500$    & $1000$   & $3800$   & $1500$  \\
    $(m_{\tilde{u}})_{33}$/GeV & $3000$   & $3000$   & $4500$    & $5000$   & $1000$   & $1500$  \\
    $(m_{\tilde{d}})_{33}$/GeV & $3000$   & $3000$   & $2000$    & $3000$   & $3000$   & $4500$  \\
    $(m_{\tilde{l}})_{33}$/GeV & $3000$   & $3000$   & $3000$    & $3000$   & $3000$   & $3000$  \\
    $(m_{\tilde{e}})_{33}$/GeV & $3000$   & $3000$   & $3000$    & $3000$   & $3000$   & $4500$  \\
    $x_t$                      & $2.2$    & $2.25$   & $-1.4$    & $-3.0$   & $3.1$    & $-1.25$ \\
    $x_b$                      & $-16.5$  & $-15.0$  & $-32.5$   & $0.0$    & $5.0$    & $38.0$  \\ \bottomrule
  \end{tabular}
  \caption{Selected \MSSM\ benchmark points inspired by the result of the parameter scan
   specified in Eq.~\eqref{eq:MSSM_parameter_scan}.}
  \label{tab:BMP_MSSM}
\end{table}

The resulting viable parameter points emerging from the scan can be divided into
groups with similar characteristics. In order to study the properties
of interesting scenarios with the \FEFTH\ calculation, we chose one
representative benchmark point from each group. The selected benchmark
points are listed below, including a brief description of their
characteristics:
\begin{itemize}
\item \textbf{(M0)} represents a standard scenario with degenerate
  \SUSY\ mass spectrum and serves as a reference for the other
  benchmark points with non-degenerate mass spectra. Parameter points
  of that kind have been thoroughly analyzed in
  Ref.~\cite{Kwasnitza:2020wli}.
\item \textbf{(M1)} is characterized by a small value of $\tan
  \beta$. During the scan, no parameter points with smaller values of
  $\tan\beta$ have been found that reproduce the experimental value of
  the Higgs mass.
\item \textbf{(M2)} was chosen due to the fact that its value of
  $|x_t|$ differs strongly from $\sqrt{6}$.
\item \textbf{(M3)} has the feature that $(m_{\tilde{q}})_{33}$ is
  significantly smaller than $(m_{\tilde{u}})_{33}$ and all other soft
  \SUSY-breaking mass parameters.
\item \textbf{(M4)} is characterized by a value of
  $(m_{\tilde{u}})_{33}$ being significantly smaller than
  $(m_{\tilde{q}})_{33}$ and all other soft \SUSY-breaking mass
  parameters.
\item \textbf{(M5)} features a small value of $(m_{\tilde{q}})_{33}$
  and $(m_{\tilde{u}})_{33}$ at the same time.
\end{itemize}

\begin{table}
  \centering
  \begin{tabular}{ccccccccc}
    \toprule
    BMP               & M0        & M1        & M2        & M3        & M4        & M5        \\ \midrule
    $M_h$/GeV         & $125.29$  & $125.37$  & $125.09$  & $124.96$  & $125.03$  & $125.02$  \\
    $\Delta M_h$/GeV  & $0.84$    & $0.87$    & $0.65$    & $0.29$    & $1.64$    & $4.32$    \\
    $M_H$/GeV         & $3041.17$ & $3069.30$ & $2115.23$ & $1425.90$ & $2919.65$ & $4395.62$ \\
    $M_A$/GeV         & $3041.22$ & $3069.16$ & $2115.50$ & $1425.96$ & $2919.76$ & $4395.04$ \\
    $M_{H^{\pm}}$/GeV & $3041.85$ & $3067.68$ & $886.72$  & $1225.21$ & $2917.66$ & $3339.8$  \\ \bottomrule
  \end{tabular}
  \caption{Higgs pole mass spectra for the \MSSM\ benchmark points
    from Table~\ref{tab:BMP_MSSM}. The estimated total uncertainty
    $\Delta M_h$ for the \SM-like Higgs boson pole mass is also
    listed.}
  \label{tab:BMP_MSSM_Spec}
\end{table}

The explicit numerical values of the parameters at the benchmark
points can be found in Table~\ref{tab:BMP_MSSM}. The Higgs mass
spectra for these scenarios, obtained with the \FEFTH\ method at 3-loop level, are given
in Table~\ref{tab:BMP_MSSM_Spec}. First of all, one should take note
of the result for (M0), which exemplifies again the findings of
the study performed in Ref.~\cite{Kwasnitza:2020wli}, where for standard
scenarios with degenerate \SUSY\ mass spectra and $x_t \lesssim 3$
theoretical uncertainties of roughly that order were ascertained.
Comparing it to the other benchmark points indicates that at least
for a considerable fraction of the scenarios from distinct
regions one finds uncertainties of about the same size. However,
for other scenarios, especially (M5), one determines a significantly
higher uncertainty of more than $4\unit{GeV}$. In this particular
case, the large uncertainty can be attributed to two sources. One
origin are the abovementioned instabilities in the calculation of the
3-loop matching corrections $\order{g_3^4 y_t^4}$ from the \Himalaya\ library,
connected to transitions between different implemented mass
hierarchies during matching scale variation. The second origin
is the huge value of $x_b$, which provokes drastic changes in
the numerical values of $(m_{\tilde{q}})_{33}$ and
$(m_{\tilde{d}})_{33}$ during renormalization group running. Together they lead to a
large variation of the calculated $M_h$ for varying $\Qm$.
With regard to these findings, it appears questionable whether
the experimental Higgs mass can be reasonably predicted
at the benchmark point (M5) with its remarkably small value of $|x_t|$.

As mentioned above, it is well-known that in particular large
$\tan \beta$ and $|x_t| \sim \sqrt{6}$ yield large values of $M_h$.
Some of the benchmark points (M1)--(M5) do not fulfill
these two conditions. Accordingly, other mecanisms are
present that compensate the negative effect on the
size of the Higgs mass. One aspect contributing positively are
clearly large values of the squark mass parameters, which increase
the stop masses and thus induce an enhancement of $M_h$.
This effect is very obvious in (M2), where $|x_t|$ differs strongly
from $\sqrt{6}$, but very large $(m_{\tilde{q}})_{33}$ and
$(m_{\tilde{u}})_{33}$ lead to a compensation of that negative
effect. Also, a very small value of either of the two squark
mass parameters can be compensated by a large value of the other
one, respectively. This interplay occurs e.g.\ in (M3) and (M4).

\section{Application to the \NMSSM}

\subsection{Model definition}

The \NMSSM\ extends the field content of the \MSSM\ by a gauge singlet
chiral superfield $\hat{S}$. In the following we consider the
$Z_3$-invariant \NMSSM\ with the superpotential
\begin{equation}
  W = Y_d \left(\hat{H}_d \cdot \hat{Q}\right) \hat{D} - Y_u \left(\hat{H}_u \cdot \hat{Q}\right) \hat{U} + Y_e \left(\hat{H}_d \cdot \hat{L}\right) \hat{E} -\lambdaNMSSM \hat{S} \left(\hat{H}_d \cdot \hat{H}_u\right) + \frac{\kappa}{3} \hat{S}^3,
\end{equation}
where $\hat{A}\cdot\hat{B}:=\epsilon_{DE}\hat{A}^D\hat{B}^E$ is the
$SU(2)$-invariant product of the two superfields $\hat{A}$ and
$\hat{B}$.  Generation and colour indices are omitted here again for
the sake of clarity. The \NMSSM-specific couplings $\lambdaNMSSM$ and
$\kappa$ are assumed to be real in order to avoid CP-violation in the
Higgs sector. The superpotential term governed by the coupling
$\lambdaNMSSM$ replaces the analogous $\mu$-term in the \MSSM. In this way,
the \NMSSM\ provides a solution to the so-called $\mu$-problem of the
\MSSM\ \cite{Polonsky:1999qd,Bae:2019dgg} by dynamically generating an
effective $\mu$-parameter if the scalar component $s$ of the
superfield $\hat{S}$ acquires a vacuum expectation value:
\begin{equation}
  \mueff = \frac{\lambdaNMSSM v_s}{\sqrt{2}} \qquad \text{with} \qquad \frac{v_s}{\sqrt{2}} = \langle s \rangle.
  \label{eq:RelMuvs}
\end{equation}
The soft-breaking Lagrangian of the \NMSSM\ is given by
\begin{align}
  \mathcal{L}_{\text{soft}} =
  & -\tilde{q}^\dagger m^2_{\tilde{q}} \tilde{q} - \tilde{u}_R^\dagger m^2_{\tilde{u}} \tilde{u}_R - \tilde{d}_R^\dagger m^2_{\tilde{d}} \tilde{d}_R - \tilde{l}^\dagger m^2_{\tilde{l}} \tilde{l} - \tilde{e}_R^\dagger m^2_{\tilde{e}} \tilde{e}_R \nonumber \\
  & -m^2_{h_d} h_d^\dagger h_d - m^2_{h_u} h_u^\dagger h_u - m^2_s |s|^2 \nonumber \\
  & - \frac{1}{2} \left( M_1 \tilde{B} \tilde{B} + M_2 \tilde{W}_a \tilde{W}_a + M_3 \tilde{g} \tilde{g} + \hc \right) \nonumber \\
  & - \Big[ (h_d \cdot \tilde{q}_L ) T_d \tilde{d}_R^* - (h_u \cdot \tilde{q}_L ) T_u \tilde{u}_R^* + (h_u \cdot \tilde{l}_L) T_e \tilde{e}_R^* \nonumber \\
  & \qquad - \lambdaNMSSM A_{\lambdaNMSSM} s \left( h_d \cdot h_u \right) + \frac{\kappa}{3} A_{\kappa} s^3 + \hc \Big],
\end{align}
where $(T_f)_{ij}=(A_f)_{ij}(Y_f)_{ij}$.  For the couplings $\lambdaNMSSM$
and $\kappa$, a perturbativity limit can be formulated at the \SUSY\
scale by demanding the absence of a Landau pole below the GUT
scale. Using the corresponding $\beta$-functions from
Ref.~\cite{Ellwanger:2009dp},
\begin{align}
  \beta_{\lambdaNMSSM} = \frac{\text{d} \lambdaNMSSM}{\text{d}(\ln Q)} & = \kappa_L \left( 4 \lambdaNMSSM^3 + 2 \lambdaNMSSM \kappa^2 + 3 \lambdaNMSSM y_t^2 - \frac{3}{5} \lambdaNMSSM g_1^2 - 3 \lambdaNMSSM g_2^2 \right) + \order{\kappa_L^2}, \\
  \beta_{\kappa} = \frac{\text{d} \kappa}{\text{d}(\ln Q)} & = \kappa_L \left( 6 \lambdaNMSSM^2 \kappa + 6 \kappa^3 \right) + \order{\kappa_L^2},
\end{align}
where $\kappa_L=1/(4\pi)^2$, that limit can be found to be approximately
\cite{Masip:1998jc}:
\begin{equation}
  \sqrt{\lambdaNMSSM^2 + \kappa^2} < 0.7.
  \label{eq:Pert_Lim}
\end{equation}
The Higgs potential resulting from the \NMSSM\ Lagrangian is given by
\begin{align}
  V_H = \ & \frac{g_1^2+g_2^2}{8}\left( h_d^\dagger h_d - h_u^\dagger h_u \right)^2 + \frac{g_2^2}{2} \left| h_d^{\dagger} h_u \right|^2 \nonumber \\
          & + \lambdaNMSSM^2 |s|^2 \left( h_u^\dagger h_u + h_d^\dagger h_d \right) + \left|\lambdaNMSSM (h_u \cdot h_d) + \kappa s^2\right|^2 \nonumber \\
          & + m_{h_d}^2 h_d^\dagger h_d + m_{h_u}^2 h_u^\dagger h_u + m_s^2 |s|^2 + \left[ \lambdaNMSSM A_{\lambdaNMSSM} s (h_u \cdot h_d) + \frac{\kappa}{3} A_{\kappa} s^3 + \hc \right].
\end{align}
The potential is required to be configured such that all Higgs fields
acquire a vacuum expectation value (VEV) by electroweak symmetry
breaking. The VEVs of the Higgs doublets $h_u$ and $h_d$ are denoted
as $v_u$ and $v_d$, respectively, and are defined as in the \MSSM,
c.f.\ Eq.~\eqref{eq:Higgs_double_VEVs}.
In this work, we restrict ourselfs to the CP-conserving \NMSSM. In
order to circumvent CP violation in the Higgs sector, all VEVs are
assumed to be real. Furthermore, all parameters appearing in the Higgs
sector are accordingly chosen to be real and the CKM matrix is set to
unity. As CP is conserved, CP-even and CP-odd neutral Higgs fields do
not mix. There is however a mixing within the CP-even and the CP-odd
Higgs sectors, respectively:
\begin{align}
  \Re(h_u^0), \Re(h_d^0), \Re(s) \ &\longrightarrow \ h_1, h_2, h_3, &
  \Im(h_u^0), \Im(h_d^0), \Im(s) \ &\longrightarrow \ A_1, A_2, G^0
\end{align}
with the CP-even mass eigenstates $h_1$, $h_2$ and $h_3$, the CP-odd
ones $A_1$ and $A_2$ as well as the Goldstone boson
$G^0$. Diagonalizing the CP-even Higgs boson mass matrix, the
tree-level mass of lightest mass eigenstate $h_1$, which is assumed to
correspond to the \SM-like Higgs boson $h$, is found to be:
\begin{equation}
  m_h^2 = m_Z^2 \cos^2(2\beta) + \lambdaNMSSM^2 v^2 \sin^2(2\beta) - \frac{\left[\sqrt{2} \lambdaNMSSM^2 v_s - \lambdaNMSSM (A_{\lambdaNMSSM} + \sqrt{2} \kappa v_s)\sin(2\beta) \right]^2 v^2}{\kappa v_s (\sqrt{2} A_{\kappa} + 4 \kappa v_s)},
  \label{eq:Mh_tree_exp}
\end{equation}
in the limit of $v \ll m_A$ and $v \ll \kappa v_s (\sqrt{2} A_{\kappa} + 4 \kappa v_s)$
(where the right side of the second inequality corresponds to the
tree-level mass of the singlet scalar), i.e.\ keeping only leading terms of
$\order{v^2}$ (see also Ref.~\cite{Bagnaschi:2022zvd}). The first
term originates from the D-terms, the second one from
F-terms and the third one from the decoupling to the singlet scalar
component. Thus the \NMSSM\ contains the following set of non-\SM\
$\DRbarPrime$ input parameters:
\begin{align}
  P_{\NMSSM}(\Qin)
  = \ & \left[P_{\MSSM}(\Qin)\setminus\left\{\mu(\Qin),B\mu(\Qin)\right\}\right] \nonumber \\
      &\cup \left\{ \lambdaNMSSM(\Qin), \kappa(\Qin), v_s(\Qin), A_{\lambdaNMSSM}(\Qin), A_{\kappa}(\Qin), m_s^2(\Qin) \right\}.
\end{align}
Furthermore, an \MSSM\ limit of the \NMSSM\ can be defined by fixing
the $\mu$ and the $B\mu$ parameters of the \MSSM\ to their effective
\NMSSM\ values
\begin{align}
  \mueff &:= \frac{\lambdaNMSSM v_s}{\sqrt{2}}, &
   B\mueff &:= \mueff\left(A_\lambdaNMSSM + \frac{\kappa v_s}{\sqrt{2}}\right),
\end{align}
and letting $\lambdaNMSSM,\kappa\to 0$. Accordingly, one has
$v_s \rightarrow \infty$ and the new scalar gauge singlet $s$ fully
decouples from the Higgs sector.

\subsection{Matching details}
\label{sec:Match_Det}

In the following we apply the \FEFTH\ method to the \NMSSM.  The
matching of the \NMSSM\ to the \SM\ is performed analogously to
Ref.~\cite{Kwasnitza:2020wli}, such that terms up to 3-loop order to
$\lambdaSM(\Qm)$ are included, c.f.\ Eq.~\eqref{eq:lambda-hat}.
At 1-loop level we include all \NMSSM\ contributions, in particular
all genuine \NMSSM\ terms that dependend on $\lambdaNMSSM$ and $\kappa$.
At 2-loop level we take the \MSSM-like terms of
$\order{g_3^2(y_t^4+y_b^4) + (y_t^2+y_b^2)^3 + (y_t^2 + y_\tau^2)^3}$
into account in the gaugeless limit, $g_1 = g_2 = 0$, and in the
zero-momentum approximation. This implies that \NMSSM-specific terms
are fully absent from $\DeltalambdaSM^{2\ell}$. At 3-loop level we
include the 3-loop \QCD\ contributions of $\order{g_3^4y_t^4}$ from
Ref.~\cite{Harlander:2018yhj}.

\subsection{Uncertainty estimation}
\label{sec:NMSSM-UncEst}

The uncertainty estimation for the predicted light CP-even Higgs pole
mass $M_h$ in our \FEFTH\ \NMSSM\ calculation is performed in an
analogous way as in the \MSSM, see Section~\ref{sec:MSSM-UncEst}. The
total uncertainty is obtained via Eq.~\eqref{eq:Comb_Uncert} from
the combination of high-scale and low-scale uncertainties. The
low-scale uncertainty is identical to the one in the MSSM, and
the high-scale uncertainty is defined via matching-scale
variation as in Eq.~\eqref{eq:DMh_Qm}. In this section, we discuss the
\NMSSM\ high-scale uncertainty. In particular, we focus on the question
to what extent the matching scale variation is a reliable method to estimate
missing \NMSSM-specific higher-order terms that depend on $\lambdaNMSSM$
and $\kappa$. The answer to this question is particularly important in
parameter regions where these couplings are of $\order{1}$. In the
following, we discuss this question in detail.

The analysis of the reliability of the matching scale variation below
is performed according to the following strategy. First, the
structure of missing \NMSSM-specific terms contributing to
$\lambdaSM(\Qm)$ up to 3-loop level is elaborated. Next, the
structure of terms simulated by matching scale variation is
derived. Since the matching scale is unphysical, the exact Higgs pole
mass $M_h^{\text{exact}}$ can not depend on it, whereas for the
calculated Higgs mass $M_h$, the cancellation of the simulated
matching scale-dependent higher-order terms is incomplete. The
non-cancelled matching scale-dependent terms are the terms simulated
by matching scale variation. Finally, the structure of the terms not
included in $\lambdaSM(\Qm)$ is compared to the structure of the
terms simulated by the matching scale variation. We will analyse to
what extent the latter incorporates the former and in this way demonstrate
the reliability of the high-scale uncertainty estimation.

\begin{figure}
  \begin{subfigure}[b]{0.34\textwidth}
    \centering
    \begin{tikzpicture}
      \pgfmathsetmacro{\R}{0.8};  
      \pgfmathsetmacro{\EL}{0.8}; 
      \draw[scalar] (-\R,0) circle (\R);
      \draw[scalar] (+\R,0) circle (\R);
      \draw[scalar] (+\R,0)++( 45:\R) -- ++ ( 45:\EL) node[above right] {$\phi$};
      \draw[scalar] (-\R,0)++(135:\R) -- ++ (135:\EL) node[above left] {$\phi$};
      \draw[scalar] (-\R,0)++(225:\R) -- ++ (225:\EL) node[below left] {$\phi$};
      \draw[scalar] (+\R,0)++(-45:\R) -- ++ (-45:\EL) node[below right] {$\phi$};
      \node[above] at (-\R,+\R) {$\phi$};
      \node[below] at (-\R,-\R) {$\phi$};
      \node[above] at (+\R,+\R) {$\phi$};
      \node[below] at (+\R,-\R) {$\phi$};
      \node[right] at (+2*\R,0) {$s$};
      \node[left]  at (-2*\R,0) {$s$};
    \end{tikzpicture}
    \caption{}
    \label{fig:mis_lambda_2l}
  \end{subfigure}
  \begin{subfigure}[b]{0.29\textwidth}
    \centering
    \begin{tikzpicture}
      \pgfmathsetmacro{\RE}{1.2}; 
      \pgfmathsetmacro{\RI}{0.4}; 
      \pgfmathsetmacro{\EL}{0.8}; 
      \draw[scalar] (0,0) circle (\RE);
      \draw[scalar] (0,0) circle (\RI);
      \draw[scalar] ( 45:\RE) -- ++( 45:\EL) node[above right] {$\phi$};
      \draw[scalar] (135:\RE) -- ++(135:\EL) node[above left ] {$\phi$};
      \draw[scalar] (225:\RE) -- ++(225:\EL) node[below left ] {$\phi$};
      \draw[scalar] (315:\RE) -- ++(315:\EL) node[below right] {$\phi$};
      \draw[scalar] (+90:\RI) -- (+90:\RE) node[left,midway] {$\phi$};
      \draw[scalar] (-90:\RI) -- (-90:\RE) node[left,midway] {$\phi$};
      \node[left]  at (  180:\RI) {$s$};
      \node[right] at (    0:\RI) {$\phi$};
      \node[right] at (    0:\RE) {$s$};
      \node[above] at ( 67.5:\RE) {$\phi$};
      \node[above] at (112.5:\RE) {$s$};
      \node[left]  at (  180:\RE) {$\phi$};
      \node[below] at (247.5:\RE) {$s$};
      \node[below] at (292.5:\RE) {$\phi$};
    \end{tikzpicture}
    \caption{}
    \label{fig:mis_lambda_3l_log}
  \end{subfigure}
  \begin{subfigure}[b]{0.34\textwidth}
    \centering
    \begin{tikzpicture}
      \pgfmathsetmacro{\R}{0.8};  
      \pgfmathsetmacro{\EL}{0.8}; 
      \draw[scalar] (-\R,0) circle (\R);
      \draw[scalar] (+\R,0) circle (\R);
      \draw[scalar] (-\R,\R) -- (-\R,-\R) node[left,midway] {$s$};
      \draw[scalar] (+\R,0)++( 45:\R) -- ++ ( 45:\EL) node[above right] {$\phi$};
      \draw[scalar] (-\R,0)++(135:\R) -- ++ (135:\EL) node[above left] {$\phi$};
      \draw[scalar] (-\R,0)++(225:\R) -- ++ (225:\EL) node[below left] {$\phi$};
      \draw[scalar] (+\R,0)++(-45:\R) -- ++ (-45:\EL) node[below right] {$\phi$};
      \node[above] at ($(-\R,0)+(   45:\R)$) {$\phi$};
      \node[above] at ($(-\R,0)+(112.5:\R)$) {$\phi$};
      \node[below] at ($(-\R,0)+(247.5:\R)$) {$\phi$};
      \node[below] at ($(-\R,0)+(  -45:\R)$) {$\phi$};
      \node[above] at (+\R,+\R) {$\phi$};
      \node[below] at (+\R,-\R) {$\phi$};
      \node[right] at (+2*\R,0) {$s$};
      \node[left]  at (-2*\R,0) {$s$};
    \end{tikzpicture}
    \caption{}
    \label{fig:mis_lambda_3l}
  \end{subfigure}
  \caption{Diagrams corresponding to contributions to
   $\lambdaSM(\Qm)$ not covered by the calculation, where
   $\phi\in\{\Re(h_u^0), \Re(h_d^0)\}$. \figurename\
   \subref{fig:mis_lambda_2l}: non-log-enhanced 2-loop contribution of
   $\order{(\lambdaNMSSM+\kappa)^4\lambdaNMSSM^2}$.  \figurename\
   \subref{fig:mis_lambda_3l_log}: log-enhanced 3-loop contribution of
   $\order{(\lambdaNMSSM+\kappa)^8}$. \figurename\
   \subref{fig:mis_lambda_3l}: non-log-enhanced 3-loop contribution of
   $\order{(\lambdaNMSSM+\kappa)^6\lambdaNMSSM^2}$.
   Recall that triple scalar couplings involving one $s$ contribute a
   vertex factor of $\order{\kappa + \lambdaNMSSM}$ and quartic vertices
   involving only $\phi$ a factor of $\order{\kappa^2/\lambdaNMSSM}$.}
  \label{fig:mis_lambda}
\end{figure}
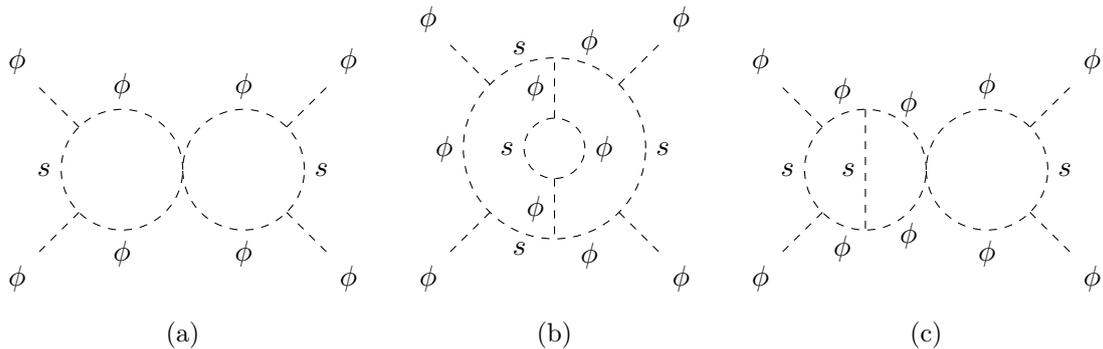

We begin by focussing on missing \NMSSM-specific corrections
of 2-loop order and will then proceed to 3-loop. In any case, these
corrections contribute to the \SM\ quartic Higgs
coupling $\lambdaSM(\Qm)$ and are eventually propagated to
$\lambdaSM(\Qpole)$ at the low scale $\Qpole$, thus affecting
$M_h^2$. Analogously, \NMSSM effects enter the other \SM\ parameters.
In this analysis, we work in the \EFT\ limit
($v\ll\MS$) which is justified as we use a hybrid approach such that the \EFT\
uncertainty can be expected to be negligibly small. Furthermore, we
assume that the tree-level expression of $\lambdaSM^\tree(\Qm)$ is of
$\order{g_{1,2}^2 + (\lambdaNMSSM+\kappa)^2}$. We omit all terms containing
the Yukawa couplings $y_b$ and
$y_\tau$ due to their small numerical impact. Also, we work in the
gaugeless limit ($g_1=g_2=0$). The only couplings remaining in this
analysis are thus $\lambdaNMSSM$, $\kappa$, $y_t$ and $g_3$.

In our calculation, the full NMSSM-specific 1-loop contributions are
included. They exhibit the following structure:
\begin{align}
  \DeltalambdaSM^{1\ell}(\Qpole)\Bigg|_{\lambdaNMSSM,\kappa}
  &= \kappa_L \Bigg[ C^{1\ell}_{\lambdaNMSSM^4} \lambdaNMSSM^4 + C^{1\ell}_{\lambdaNMSSM^3 \kappa} \lambdaNMSSM^3 \kappa + \cdots + C^{1\ell}_{\kappa^5/\lambdaNMSSM} \frac{\kappa^5}{\lambdaNMSSM} + C^{1\ell}_{\kappa^{6}/\lambdaNMSSM^2} \frac{\kappa^{6}}{\lambdaNMSSM^2} \nonumber \\
 &\phantom{= \kappa_L \Big(} + y_t^2 \Bigg( C^{1\ell}_{y_t^2 \lambdaNMSSM^2} \lambdaNMSSM^2 + C^{1\ell}_{y_t^2 \lambdaNMSSM \kappa} \lambdaNMSSM \kappa + C^{1\ell}_{y_t^2 \kappa^2} \kappa^2 \Bigg) \Bigg],
\label{eq:Delta_lambda_2l_NMSSM_spec}
\end{align}
where all $\DRbarPrime$ parameters are expressed at the scale $\Qm$.
The coefficients $C^{n\ell}_{\cdots}$ and $\tilde{C}^{n\ell}_{\cdots}$
appearing in Eq.~\eqref{eq:Delta_lambda_2l_NMSSM_spec} and also hereafter are
of $\order{1}$, by assuming the ratios $\mueff/\MS$, $A_{\lambdaNMSSM}/\MS$
and $A_{\kappa}/\MS$ to be of the same order. It is noteworthy that
$\lambdaNMSSM$ also appears in the denominator due to Higgs couplings
proportional to $\kappa v_s$, where $v_s$ is reexpressed in terms of
$\lambdaNMSSM$ and $\mueff$ as $v_s=\sqrt{2}\mueff/\lambdaNMSSM$, c.f.\
Eq.~\eqref{eq:RelMuvs}. By performing renormalization group running in the
\SM from $\Qm$ down to $\Qlow$, all terms of leading and subleading
order in large logarithms are fully resummed. Accordingly, at 2-loop
level only \NMSSM-specific terms that are free of large logarithms are
missing in our calculation, corresponding to diagrams like the one given in
Figure~\ref{fig:mis_lambda_2l}. These non-logarithmic 2-loop contributions
are of the form
\begin{align}
  \DeltalambdaSM^{2\ell}(\Qpole)\Bigg|^{\miss}_{\lambdaNMSSM,\kappa}
  &= \kappa_L^2 \Bigg[ C^{2\ell}_{\lambdaNMSSM^6} \lambdaNMSSM^6 + C^{2\ell}_{\kappa \lambdaNMSSM^5} \lambdaNMSSM^5 \kappa + \cdots + C^{2\ell}_{\kappa^9/\lambdaNMSSM^3} \frac{\kappa^9}{\lambdaNMSSM^3} + C^{2\ell}_{\kappa^{10}/\lambdaNMSSM^4} \frac{\kappa^{10}}{\lambdaNMSSM^4} \nonumber \\
  &\phantom{= \kappa_L^2 \Bigg[} + y_t^2 \left( C^{2\ell}_{y_t^2 \lambdaNMSSM^4} \lambdaNMSSM^4 + C^{2\ell}_{y_t^2 \lambdaNMSSM^3 \kappa} \lambdaNMSSM^3 \kappa + \cdots + C^{2\ell}_{y_t^2 \kappa^6/\lambdaNMSSM^2} \frac{\kappa^6}{\lambdaNMSSM^2} \right) \nonumber \\
  &\phantom{= \kappa_L^2 \Bigg[} + y_t^2 g_3^2 \left( C^{2\ell}_{y_t^2 g_3^2 \lambdaNMSSM^2} \lambdaNMSSM^2 + C^{2\ell}_{y_t^2 g_3^2 \lambdaNMSSM \kappa} \lambdaNMSSM \kappa + C^{2\ell}_{y_t^2 g_3^2 \kappa^2} \kappa^2 \right)  \nonumber\\
  &\phantom{= \kappa_L^2 \Bigg[} + y_t^4 \left( C^{2\ell}_{y_t^4 \lambdaNMSSM^2} \lambdaNMSSM^2 + C^{2\ell}_{y_t^4 \lambdaNMSSM \kappa} \lambdaNMSSM \kappa + C^{2\ell}_{y_t^4 \kappa^2} \kappa^2 \right) \Bigg].
\end{align}
At 3-loop level, the missing contributions include terms of first
order in large logarithms, e.g.\ corresponding to diagrams shown in
Figure~\ref{fig:mis_lambda_3l_log}, and non-logarithmic terms, e.g.\
corresponding to diagrams shown in Figure~\ref{fig:mis_lambda_3l}.
These missing terms are thus of the form
\begin{align}
  \DeltalambdaSM^{3\ell}(\Qpole)\Bigg|^{\miss}_{\lambdaNMSSM,\kappa}
  &= \kappa_L^3 \Bigg[ \left( \tilde{C}^{3\ell}_{\lambdaNMSSM^8} \lambdaNMSSM^8 + \cdots + \tilde{C}^{3\ell}_{\kappa^{14}/\lambdaNMSSM^6} \frac{\kappa^{14}}{\lambdaNMSSM^6} \right) + y_t^2 \left( \tilde{C}^{3\ell}_{y_t^2 \lambdaNMSSM^6} \lambdaNMSSM^6 + \cdots \right) \nonumber \\
  &\phantom{= \kappa_L^3 \Big(} + y_t^2 g_3^2 \left( \tilde{C}^{3\ell}_{y_t^2 g_3^2 \lambdaNMSSM^4} \lambdaNMSSM^4 + \cdots \right) + y_t^4 \left( \tilde{C}^{3\ell}_{y_t^4 \lambdaNMSSM^4} \lambdaNMSSM^4 + \cdots \right) \nonumber \\
  &\phantom{= \kappa_L^3 \Big(} + y_t^2 g_3^4 \left( \tilde{C}^{3\ell}_{y_t^2 g_3^4 \lambdaNMSSM^2} \lambdaNMSSM^2 + \cdots \right) + y_t^4 g_3^2 \left( \tilde{C}^{3\ell}_{y_t^4 g_3^2 \lambdaNMSSM^2} \lambdaNMSSM^2 + \cdots \right) \nonumber \\
  &\phantom{= \kappa_L^3 \Big(} + y_t^6 \left( \tilde{C}^{3\ell}_{y_t^6 \lambdaNMSSM^2} \lambdaNMSSM^2 + \cdots \right) \Bigg] L + \text{non-log-enhanced terms},
\end{align}
where $L=\ln(\Qpole/\Qm)$ represents a large logarithm and terms
not enhanced by a large logarithm contain the same couplings structure as
the logarithmically enhanced ones. The coeffcients of the latter are
denoted by $\tilde{C}^{n\ell}_{\cdots}$ to distinguish them from the
coefficients $C^{n\ell}_{\cdots}$ of the former ones.

In order to thoroughly work out the structure of the terms simulated
by the variation of the matching scale, we shortly recall some details
of the procedure. We start this consideration with $M_h^2$ as the
expression to be calculated in the \SM\ using the $\MSbar$ scheme at
the renormalization scale $\Qpole$,
\begin{equation}
  M_h^2 = \lambdaSM \hat{v}^2 + \Delta M_h^{2,1\ell,\SM} + \Delta M_h^{2,2\ell,\SM} + \cdots.
  \label{eq:MhCalc}
\end{equation}
The $n$-loop \SM\ contributions $\Delta M_h^{2,n\ell,\SM}$ to $M_h^2$
depend on the \SM\ $\MSbar$ parameters $\hat{P}_\SM(\Qpole)$ from
Eq.~\eqref{eq:Phat_SM}, all given at the scale $\Qpole$. Any \SM\
parameter $\hat{p}\in\hat{P}_\SM(\Qpole)$ given at that
scale\footnote{In the following we omit the renormalization scale from
  the \SM\ $\MSbar$ parameters $\hat{p}\in\hat{P}_\SM(\Qpole)$, if
  they are given at the scale $\Qpole$.  We explicitely specify the
  scale, if it differs from $\Qpole$.} is obtained by performing
renormalization group running down from the matching scale $\Qm$. The
relation between a parameter $\hat{p}=\hat{p}(\Qpole)$ and its value
$\hat{p}(\Qm)$ arises from the integration of the $\beta$ function
$\beta_{\hat{p}}$ via
\begin{equation}
  \hat{p} = \hat{p}(\Qm) - \int_{\ln \Qpole}^{\ln \Qm} \beta_{\hat{p}}(Q)~\text{d}(\ln Q).
  \label{eq:ParRG}
\end{equation}
The \SM\ parameters
at the renormalization scale $\Qm$, expressed in terms of \NMSSM\ parameters
$p\in P(\Qm)$ at that scale\footnote{In the following we omit the
renormalization scale from the \NMSSM\ $\DRbarPrime$ parameters
$p\in P(\Qm)$, if they are given at the scale $\Qm$.  We explicitely
specify the scale, if it differs from $\Qm$.}, are obtained by using the matching
relations of the form
\begin{equation}
  \hat{p}(\Qm) = \hat{p}^\tree(p) + \Delta p^{1\ell}(p,\Qm) + \cdots.
  \label{eq:MatCond}
\end{equation}
In this way, the value of $\hat{p}$ features an implicit dependence on
the choice of $\Qm$. However, since the matching scale is an
unphysical scale, $\hat{p}$ would not vary under changes in $\Qm$ in
an exact all-order calculation. In the \FEFTH\ calculation, a
dependence on $\Qm$ arises due to the truncation at a specific loop
order. The following derivative is then a measure for the
arising variation:
\begin{align}
  \frac{\text{d}\hat{p}}{\text{d}(\ln \Qm)}
  &\overset{\text{\eqref{eq:ParRG}}}{=} \frac{\text{d} \hat{p}(\Qm)}{\text{d}(\ln \Qm)} - \beta_{\hat{p}}(\Qm) \\
  &\overset{\text{\eqref{eq:MatCond}}}{=} \sum_{p} \frac{\partial \hat{p}^{\tree}}{\partial p} \beta_p(\Qm) + \sum_{k\geq1} \left( \sum_{p} \frac{\partial \Delta p^{k\ell}}{\partial p} \beta_p(\Qm) + \frac{\partial \Delta p^{k\ell}}{\partial (\ln \Qm)} \right) - \beta_{\hat{p}}(\Qm).
    \label{eq:ParVar}
\end{align}
The squared pole mass $M_h^2$, calculated by a truncated perturbation
series, also depends upon $\Qm$, because it is obtained from the
$\Qm$-dependent \SM\ parameters.  Via Eq.~\eqref{eq:MhCalc}, an
expression for the derivative of $M_h^2$ with respect to the matching
scale can be found, connecting it to Eq.~\eqref{eq:ParVar}:
\begin{equation}
  \frac{\text{d}M_h^2}{\text{d}(\ln \Qm)}
  = \sum_{\hat{p}} \frac{\partial}{\partial \hat{p}} \left( \lambdaSM \hat{v}^2 + \sum_{n\geq1} \Delta M_h^{2,n\ell,\SM} \right) \frac{\text{d} \hat{p}}{\text{d}(\ln \Qm)}.
  \label{eq:MatVar}
\end{equation}
The expression in Eq.~\eqref{eq:MatVar} is required to vanish when all
orders are taken into account, as the physical Higgs pole mass must
not depend on the matching scale.

In order for the matching scale variation to be an appropriate
uncertainty estimation method, we demand the method to simulate all
missing terms depending on $\lambdaNMSSM$ and $\kappa$ and therefore
constituting \NMSSM -specific terms, as mentioned above.
To see whether this is indeed the case, we can explicitely evaluate
Eq.~\eqref{eq:MatVar} at the respective order in loops and couplings.
At 2-loop level, we find in general at $\order{\mathcal{S}}$, with
$\mathcal{S}$ representing any relevant coupling structure:
\begin{align}
  \frac{\text{d}M_h^2}{\text{d}(\ln \Qm)} \Bigg|_{\mathcal{S}}
  &= \frac{\partial}{\partial \lambdaSM} \left( \lambdaSM \hat{v}^2 + \order{\kappa_L} \right) \Bigg[ \sum_p \left( \frac{\partial \lambdaSM^\tree}{\partial p} \beta_{p}(\Qm) + \frac{\partial \DeltalambdaSM^{1\ell}}{\partial p} \beta_p(\Qm) \right) \nonumber \\
  &\phantom{=}{}\qquad + \frac{\partial \DeltalambdaSM^{1\ell}}{\partial (\ln \Qm)} + \underbrace{\frac{\partial \DeltalambdaSM^{2\ell}}{\partial (\ln \Qm)}}_{\text{incomplete}} - \beta_{\lambdaSM}(\Qm) + \order{\kappa_L^3} \Bigg] \Bigg|_{\mathcal{S}} \nonumber \\
  &\phantom{=}{} + \frac{\partial}{\partial \hat{v}} \left( \lambdaSM \hat{v}^2 + \order{\kappa_L} \right) \Bigg[ \sum_p \left( \frac{\partial \hat{v}^\tree}{\partial p} \beta_{p}(\Qm) + \frac{\partial \Delta v^{1\ell}}{\partial p} \beta_p(\Qm) \right) \nonumber \\
  &\phantom{=}{}\qquad + \frac{\partial \Delta v^{1\ell}}{\partial (\ln \Qm)} + \underbrace{\frac{\partial \Delta v^{2\ell}}{\partial (\ln \Qm)}}_{\text{incomplete}} - \beta_{\hat{v}}(\Qm) + \order{\kappa_L^3} \Bigg] \Bigg|_{\mathcal{S}}
    \label{eq:MatVar2Loop_1} \\
  &= - \left. \left( v^2 \frac{\partial \DeltalambdaSM^{2\ell}}{\partial (\ln \Qm)} \Bigg|^{\miss} + 2 \lambdaSM^\tree v \frac{\partial \Delta v^{2\ell}}{\partial (\ln \Qm)} \Bigg|^{\miss} \right) \right|_{\mathcal{S}}.
  \label{eq:MatVar2Loop_2}
\end{align}
The terms above containing partial derivatives of $\DeltalambdaSM^{2\ell}$
and $\Delta v^{2\ell}$ with respect to $\ln \Qm$ are incomplete in the
performed \FEFTH\ calculation as their NMSSM-specific parts are
omitted. However, their full inclusion would be necessary in order to
ensure the complete cancellation of $\Qm$-dependent terms at
$\order{\mathcal{S}}$. The change in the Higgs mass under matching scale
variation at that order therefore corresponds to the absent terms
(denoted by ``miss'') in the way indicated in Eq.~\eqref{eq:MatVar2Loop_2}.
Also, one can read off from Eq.~\eqref{eq:MatVar2Loop_1} which present
terms in the calculation of $M_h^2$ are the origin for simulated
terms of $\order{\mathcal{S}}$. These stem from the tree-level
contribution as well as the 1-loop corrections to $M_h^2$ in the \SM,
incorporating logarithmic as well as non-logarithmic terms.

Eq.~\eqref{eq:MatVar2Loop_2} demonstrates that the logarithmic parts
of the 2-loop matching correction to $\lambdaSM$ and $\hat{v}$ have
to be analysed in order to clarify whether a specific needed coupling
structure is generated during matching scale variation. This is the
case if for the regarded order in the couplings, a matching
contribution exists which also contains a logarithm of $\Qm$.
We can elucidate that further by regarding the concrete example of the
2-loop contributions of $\order{\lambdaNMSSM^6 v^2}$ to $M_h^2$. At that order,
we yield:
\begin{equation}
  \frac{\text{d}M_h^2}{\text{d}(\ln \Qm)} \Bigg|_{\lambdaNMSSM^6 v^2} = - v^2 \frac{\partial \DeltalambdaSM^{2\ell}}{\partial (\ln \Qm)} \Bigg|_{\lambdaNMSSM^6}^{\miss} - 2 \lambdaSM^\tree\Big|_{\lambdaNMSSM^2} v \frac{\partial \Delta v^{2\ell}}{\partial (\ln \Qm)} \Bigg|_{\lambdaNMSSM^4 v}^{\miss}.
  \label{eq:MatVarLam6_2}
\end{equation}
Since there are UV-divergent \NMSSM -specific 2-loop corrections
of $\order{\lambdaNMSSM^6}$ to the quartic self-interaction vertex of
the \SM -like Higgs, one can infer that logarithmic contributions
to the 2-loop matching relation for $\lambdaSM$ enter at that
order. With that, we can conclude that the maching
scale variation generates 2-loop terms of $\order{\lambdaNMSSM^6 v^2}$
in $M_h^2$, as they have a counterpart in the explicitely
$\Qm$-dependent part of the 2-loop maching corrections
$\DeltalambdaSM^{2\ell}$ and $\Delta v^{2\ell}$. Similar
analyses for all other appearing \NMSSM -specific coupling structures
at 2-loop level lead to the same positive result. The matching scale variation
can thus be deemed an appropriate method to estimate the uncertainty of the
2-loop \FEFTH\ calculation w.r.t.\ the missing \NMSSM -specific terms.

At 3-loop level, a similar analysis can be applied. Here, we find an
analogous result as at 2-loop level. In particular, missing terms enhanced by
large logarithms $L$ are generated by matching scale variation. This can be
exemplified at $\order{\lambdaNMSSM^8 v^2 L}$ evaluating again
Eq.~\eqref{eq:MatVar} for this specific case:
\begin{align}
  \frac{\text{d}M_h^2}{\text{d}(\ln \Qm)} \Bigg|_{\lambdaNMSSM^8 v^2 L}
  &= -\hat{v}^2\Big|_{\lambdaNMSSM^2 v^2 L} \frac{\partial \DeltalambdaSM^{2\ell}}{\partial (\ln \Qm)} \Bigg|_{\lambdaNMSSM^6}^{\miss}
    - 2 \lambdaSM^\tree\Big|_{\lambdaNMSSM^4 L} v \frac{\partial \Delta v^{2\ell}}{\partial (\ln \Qm)} \Bigg|_{\lambdaNMSSM^4 v}^{\miss} \nonumber \\
  &\phantom{=}{} - 2 \lambdaSM^\tree\Big|_{\lambdaNMSSM^2} \hat{v}\Big|_{\lambdaNMSSM^2 L} \frac{\partial \Delta v^{2\ell}}{\partial (\ln \Qm)} \Bigg|_{\lambdaNMSSM^4 v}^{\miss}.
  \label{eq:l8v2L-terms}
\end{align}
Here, the large logarithm $L$ appears due to the running of the \SM\
parameters from $\Qm$ to $\Qpole$ by applying Eq.~\eqref{eq:ParRG}.
The appearance of the large logarithm $L$ corresponds to the fact that
$\NNLL$ terms of $\order{\lambdaNMSSM^8v^2}$ are not included in our
calculation. The matching scale variation estimates these missing
terms, though. Similarly, $\NNLL$ terms of higher orders are covered as well.

From the outcomes of the presented analysis above, we infer that matching
scale variation serves as a reliable mean for estimating the high-scale
uncertainty of the calculated Higgs mass $M_h^2$. Especially, this is also
true for the \NMSSM -specific corrections entering the matching relations
between the \SM and the \NMSSM.

\subsection{Numerical results}

In the following we present numerical results for the 3-loop \FEFTH\
calculation of the light CP-even Higgs boson pole mass $M_h$ in
\NMSSM. In particular we study the impact of the \NMSSM-specific
couplings $\lambdaNMSSM$ and $\kappa$ on the predicted value of $M_h$. We
furthermore inspect whether the $\lambdaNMSSM$- and $\kappa$-dependent
uncertainties predominate against \MSSM-like ones.

In our numerical study we set the \NMSSM\ $\DRbarPrime$ parameters to
the values given in
Eqs.~\eqref{eq:parameter_condition_MSSM_mf}--\eqref{eq:parameter_condition_MSSM_Af},
if not stated otherwise.  The remaining \NMSSM\ parameters, in
particular $\lambdaNMSSM(\Qin)$, $\kappa(\Qin)$, $\mueff(\Qin)$,
$A_{\lambdaNMSSM}(\Qin)$ and $A_{\kappa}(\Qin)$ are set to various values,
depending on which specific kind of scenario is of interest.

\subsubsection{Validation}

\begin{figure}
  \begin{subfigure}[b]{0.5\textwidth}
    \includegraphics[width=\textwidth]{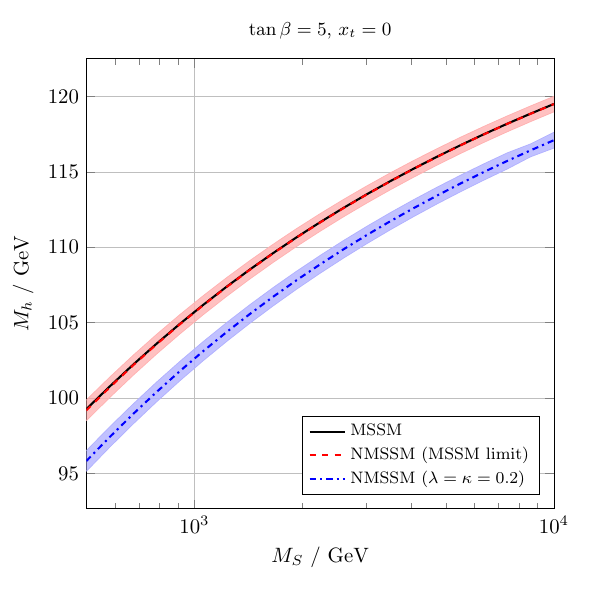}
  \end{subfigure}
  \begin{subfigure}[b]{0.5\textwidth}
    \includegraphics[width=\textwidth]{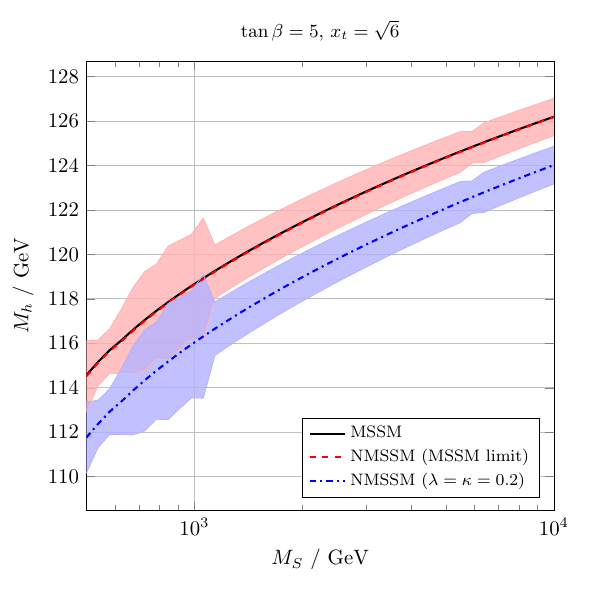}
  \end{subfigure}
  \caption{Comparison between the calculated Higgs mass values in the
    \MSSM\ and the \NMSSM\ in dependence on {\MS}, obtained from the
    \FEFTH\ approach implemented in \FS. The left panel depicts the
    curves for $x_t=0$ and the right one for $x_t=\sqrt{6}$. The red
    band represents the estimated uncertainty for the \NMSSM\
    calculation in the \MSSM\ limit, while the blue band indicates the
    uncertainty for $\lambdaNMSSM=\kappa=0.2$.}
  \label{fig:Mh_NMSSM_MSSM}
\end{figure}

\begin{figure}
  \begin{subfigure}[b]{0.5\textwidth}
    \includegraphics[width=\textwidth]{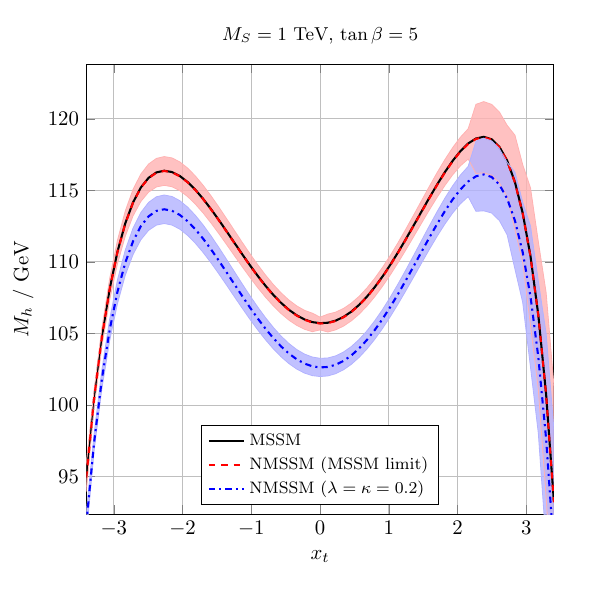}
  \end{subfigure}
  \begin{subfigure}[b]{0.5\textwidth}
    \includegraphics[width=\textwidth]{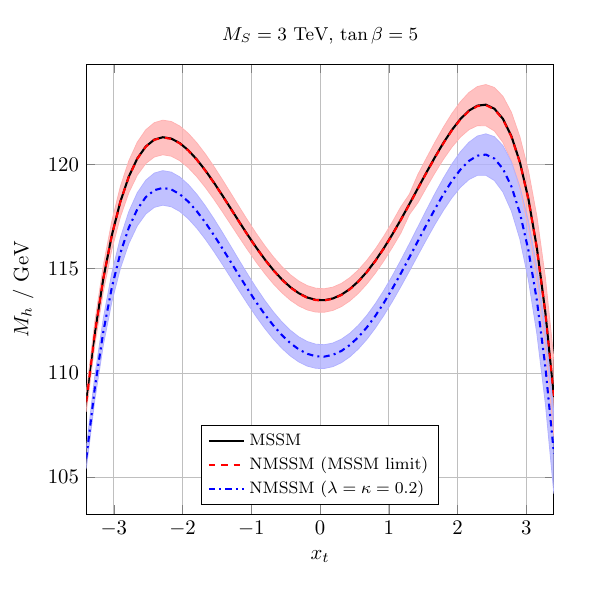}
  \end{subfigure}
  \caption{Comparison between of the \FEFTH\ Higgs mass calculation in
    the \MSSM\ and the \NMSSM\ in dependence on $x_t$, in analogy to
    Figure~\ref{fig:Mh_NMSSM_MSSM}. The left panel depicts the curves for
    $\MS=1\unit{TeV}$ and the right one for $\MS=3\unit{TeV}$.}
  \label{fig:Mh_NMSSM_MSSM_xt}
\end{figure}

In order to validate the newly implemented \FEFTH\ calculation in
the \NMSSM, we compare the obtained output with already existing
results. These incorporate fixed-order (FO) and \EFT\ calculations as well as
\FEFTH\ results for the \MSSM\ in \FS.

As a first test, the Higgs pole mass calculated in the \MSSM\
limit of the \NMSSM\ may be compared to the actual \MSSM\ result based on the \FEFTH\
approach. For that comparison, we regard scenarios with degenerate
SUSY mass parameters set to the common scale $\MS$. For the {\NMSSM},
however, we make an exception concerning the mass-dimension parameters
$A_{\lambdaNMSSM}$ and $A_{\kappa}$ in order to achieve accordance to the
corresponding \MSSM\ parameter point: $A_{\lambdaNMSSM}$ is chosen such that
$m_A=\MS$ is fulfilled. $A_{\kappa}$ is then set to $A_{\kappa}=-|A_{\lambdaNMSSM}|$.
In Figure~\ref{fig:Mh_NMSSM_MSSM}, the two curves corresponding to the
\MSSM\ Higgs mass and the \NMSSM\ Higgs mass in the \MSSM\ limit
are given in both subfigures in dependence on $\MS$, clearly indicating
a very good agreement. Also the estimated uncertainties indicate a
stable calculation for $0.5\unit{TeV} \leq \MS \leq 10\unit{TeV}$, merely
for $x_t=\sqrt{6}$ and $\MS \sim 1\unit{TeV}$ jumps appear in the
uncertainty caused by hierarchy transitions in \Himalaya.
A very similar picture arises in Figure~\ref{fig:Mh_NMSSM_MSSM_xt}
where a very good agreement between the \MSSM\ result and the \NMSSM\
result in the \MSSM\ limit was achieved and a nearly constant
uncertainty was found, except for several in stabilities for
$x_t \sim \sqrt{6}$ and for small $\MS$.

Next, a comparison with other \NMSSM\ Higgs mass calculations also
available in \FS\ is performed. The \FEFTH\ calculation, as it is now
implemented in the model \texttt{NMSSMEFTHiggs}, allows to switch on
and off the accessable 1-, 2- and 3-loop corrections to the
$\lambdaSM$-matching, as described in
Section~\ref{sec:Match_Det}. These three options are denoted below
as ``FEFT $n\ell$'', where $n$ represents the highest loop-order of the
corrections taken into account. Besides the hybrid \FEFTH\ calculation,
also a fixed-order calculation in the \NMSSM\ is available within \FS. That
calculative approach has been implemented in the recently created
model \texttt{NUHNMSSMHimalaya}. Similarly to the \FEFTH\ calculation,
it allows to include \MSSM-like 2-loop corrections of
$\order{g_3^2(y_t^4+y_b^4) + (y_t^2+y_b^2)^3 + (y_t^2 + y_\tau^2)^3}$
to the Higgs mass as well as {\MSSM}-like 3-loop corrections of
$\order{g_3^4y_t^4}$. If the latter ones are taken into account, it is
denoted here as ``FO $3\ell$'', otherwise it is referred to as ``FO
$2\ell$''.

\begin{figure}
  \begin{subfigure}[b]{0.5\textwidth}
    \includegraphics[width=\textwidth]{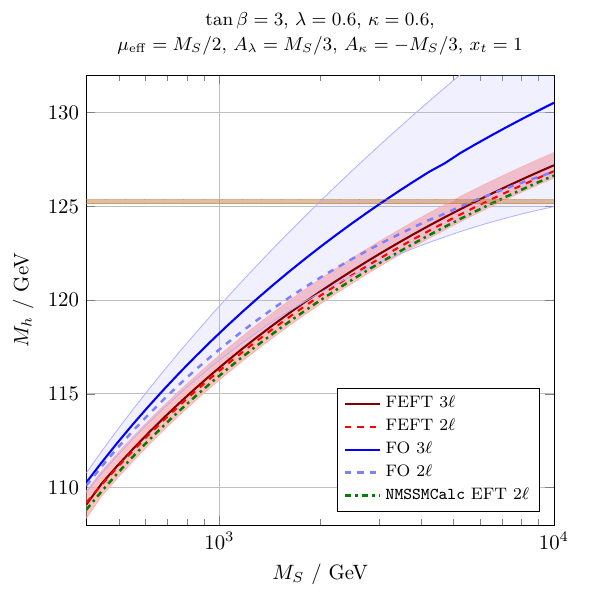}
  \end{subfigure}
  \begin{subfigure}[b]{0.5\textwidth}
    \includegraphics[width=\textwidth]{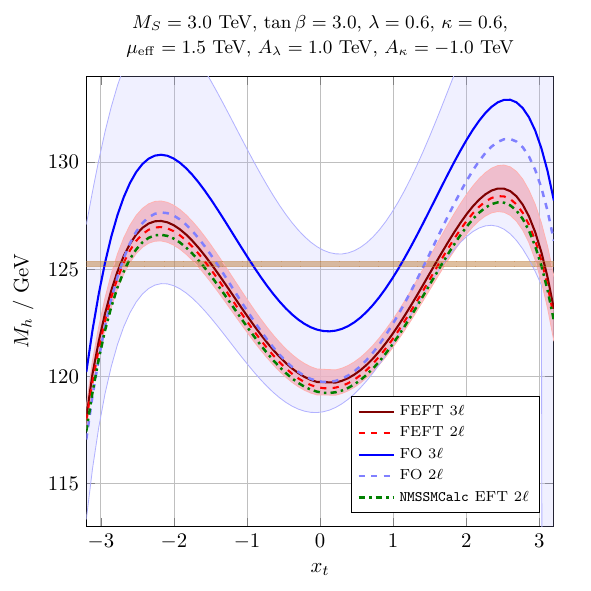}
  \end{subfigure}
  \caption{Comparison between the calculated Higgs mass values obtained
    from different approaches implemented in \FS. The
    dependence $M_h(\MS)$ is shown on the left side, whereas $M_h(x_t)$
    is shown on the right side. The red band represents the estimated
    uncertainty for the 3-loop \FEFTH calculation, while the blue band
    indicates the uncertainty for the 3-loop FO calculation. Additionally,
    the \NMSSMCalc\ result obtained with pole mass matching including
    full 1-loop matching corrections and 2-loop corrections of
    $\mathcal{O}(\hat{g}_3^2 \hat{y}_t^4 + \hat{y}_t^6)$ is plotted.}
  \label{fig:Mh_EFT_FO_Comp}
\end{figure}

In the left plot of Figure~\ref{fig:Mh_EFT_FO_Comp} we compare the
calculated Higgs pole mass in dependence on $\MS$ obtained by the \FEFTH
approach with the FO 2-loop and 3-loop results. There is a relatively
good agreement between the four corresponding curves for
$\MS < 1\unit{TeV}$ as one would expect it for a comparison between
hybrid and FO results. However, there is still a deviation of
$\sim 1\unit{GeV}$ for very low values of $\MS$, which originate from
a different inclusion of terms of higher order between the two
calculations. For larger \SUSY\ scales,
there are more significant deviations of the order $\sim 5\unit{GeV}$, which
can be explained by numerically important, log-enhanced terms of
higher order, which are not included in the \FO calculation at 2-loop
and 3-loop level, respectively, but taken into account in the \FEFTH
calculation. The broadening of the corresponding uncertainty band for
increasing $\MS$ mirrors this fact. However, the uncertainty for the
\FEFTH\-result remains nearly constant in the whole regarded scale
range. Furthermore, the difference between the 2-loop and 3-loop \FEFTH
results are remarkably small. These outcomes strongly suggest
excellent convergence properties of the \FEFTH\ method.

In the right panel of Figure~\ref{fig:Mh_EFT_FO_Comp}, the dependence
of $M_h$ on $x_t$ is shown. There, the differences between the results
of the \FEFTH- and the \FO-calculation are evident as well. Very
remarkable is the significantly smaller width of the uncertainty band
obtained with the \FEFTH calculation, in accordance with the findings
above. \FEFTH- and 3-loop \FO-results are separated by a gap of about
$2\unit{GeV}$, which is nearly independent of $x_t$. Also these observations
indicate a much higher reliability of the FEFT-calculation.

Furthermore, a numerical comparison was carried out with the results from
\NMSSMCalc\ \cite{Baglio:2013iia,Dao:2019qaz,Dao:2021khm,Borschensky:2024utz},
which is based on a similar calculation including a hybrid approach with the
option of choosing a Higgs pole mass matching. The corresponding result,
obtained using pole mass matching with full 1-loop matching corrections and
2-loop corrections of $\mathcal{O}(\hat{g}_3^2 \hat{y}_t^4 + \hat{y}_t^6)$, is also shown in
Figure~\ref{fig:Mh_EFT_FO_Comp}. The comparsion reveals a very good agreement between
the two codes in general. The major difference to the 2-loop version of \FEFTH\
originates from the additional inclusion of 3-loop corrections in the relation
between the top Yukawa coupling $\hat{y}_t$ and the top pole mass $M_t$ which
implies a nearly constant shift in $M_h$ of about $0.5\unit{GeV}$ under variation
of $\MS$ and $x_t$, as Figure~\ref{fig:Mh_EFT_FO_Comp} illustrates.\footnote{In
order to trace back further origins of the difference, we internally applied
a modification to the \NMSSMCalc\ code to omit the 3-loop corrections to the top
pole mass thanks to the support by Martin Gabelmann. The Higgs mass prediction
obtained by the modified code is significantly closer to the \FS\ results. The
difference is of the order of $0.3\unit{GeV}$.} 
Further differences arise from the additional matching corrections involving
the Yukawa couplings $y_b$ and $y_{\tau}$ taken into account in
\FS\ as well as from different choices of the parametrisation. While
\NMSSMCalc\ uses \EFT parametrization, \FS uses full-model parametrization, leading
to different higher-order terms in the high-scale matching  including an effective
resummation of terms involving $x_t$, see the discussion in Sec.~2 and
Refs.~\cite{Kwasnitza:2020wli,Kwasnitza:2021idg}. Furthermore, \NMSSMCalc differs
in details of the low-scale matching between $\MSbar$ parameters and observables.
Particularly in the algorithm for extracting the value of the top Yukawa coupling
$\hat{y}_t$ from the top pole mass, different higher-order corrections are taken
into account beyond the nominal accuracy.

Also, the results from Ref.~\cite{Bagnaschi:2022zvd} were compared to the \FS\ predictions.
There, an \EFT\ approach was applied including \NMSSM-specific 2-loop QCD corrections
to the matching. The comparison indicates a very good agreement as well. Additionally,
it is worth mentioning that the study performed in Ref.~\cite{Bagnaschi:2022zvd}
led to the conclusion that \NMSSM-specific matching corrections contribute substantially
only for quite large values of $\lambda$. This finding further corroborates the
approach implemented in \FS.

\subsubsection{\NMSSM-specific effects}

In this subsection, the behaviour and numerical impact of \NMSSM-specific
contributions to the \SM-like Higgs pole mass $M_h$ is analyzed. In
particular, missing higher order terms of that kind are estimated based
on the procedure presented in Section~\ref{sec:NMSSM-UncEst}. They will also be
compared to missing \MSSM-like terms.

\begin{figure}
  \begin{subfigure}[b]{0.5\textwidth}
    \includegraphics[width=\textwidth]{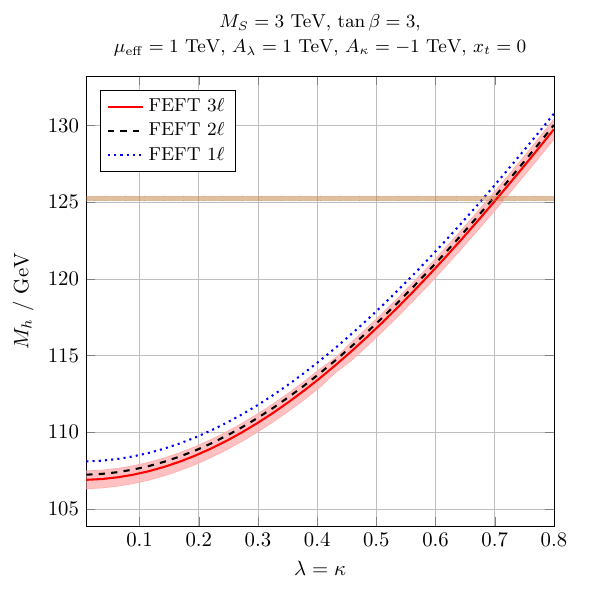}
  \end{subfigure}
  \begin{subfigure}[b]{0.5\textwidth}
    \includegraphics[width=\textwidth]{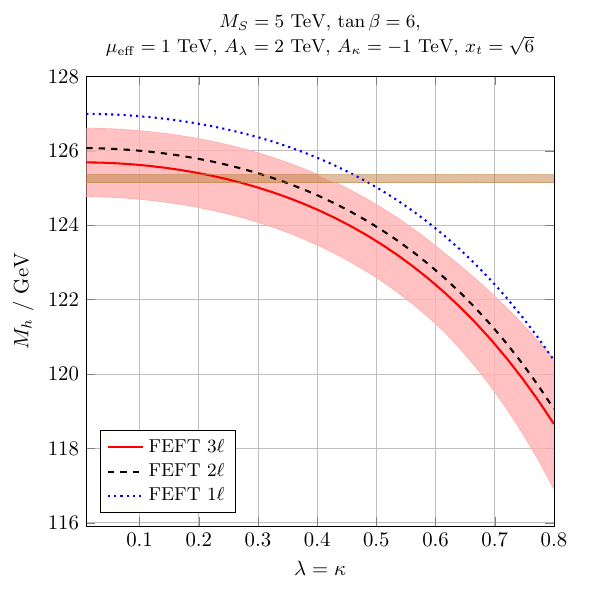}
  \end{subfigure}
  \caption{Prediction of the Higgs pole mass $M_h$ as calculated by
    the \FEFTH\ (FEFT) method at 1-, 2- and 3-loop precision in
    dependence on $\lambdaNMSSM=\kappa$ for different scenrios. The red
    curve represents the \FEFTH $3\ell$ result, for which the red band
    indicates the estimated uncertainty. The horizontal brownish band
    marks again the experimentally determined value of $M_h$ with its
    $1\sigma$-uncertainty.}
  \label{fig:Mh_lambda_kappa_1}
\end{figure}

Figure~\ref{fig:Mh_lambda_kappa_1} illustrates interplay of
$\tan \beta$ and Yukawa couplings $\lambdaNMSSM$ and $\kappa$ in affecting
Higgs pole mass $M_h$. The qualitative behaviour can be understood by
considering the well-known dependence of the tree-level contribution
on these two couplings. For large $\tan \beta$, the second term on the
right side of Eq.~\eqref{eq:Mh_tree_exp} vanishes, while the
non-vanishing third term leads to a negative shift which increases
with growing $\lambdaNMSSM=\kappa$. For small $\tan \beta$ however, the
second term dominates compared to the third one such that $M_h$
increases with $\lambdaNMSSM$. Figure~\ref{fig:Mh_lambda_kappa_1} thus
demonstrates again, how strongly effects depending on $\lambdaNMSSM$ and
$\kappa$ can influence $M_h$ and accordingly their ability to
alleviate the fine-tuning problem with regard to $x_t$ in the \MSSM.

\begin{figure}
  \begin{subfigure}[b]{0.5\textwidth}
    \includegraphics[width=\textwidth]{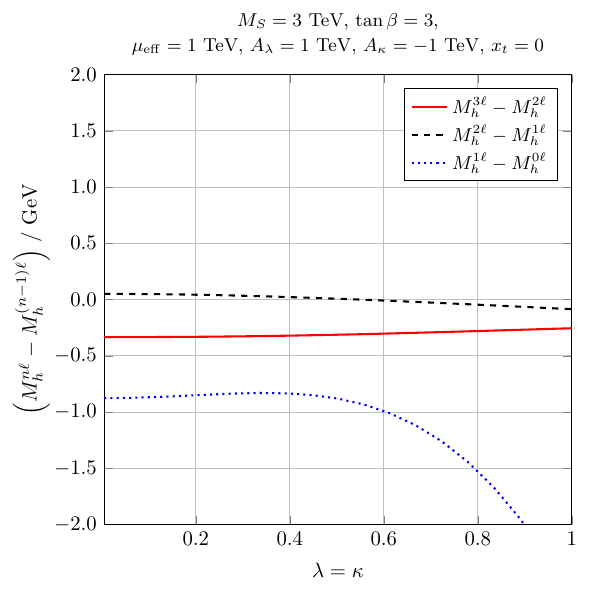}
  \end{subfigure}
  \begin{subfigure}[b]{0.5\textwidth}
    \includegraphics[width=\textwidth]{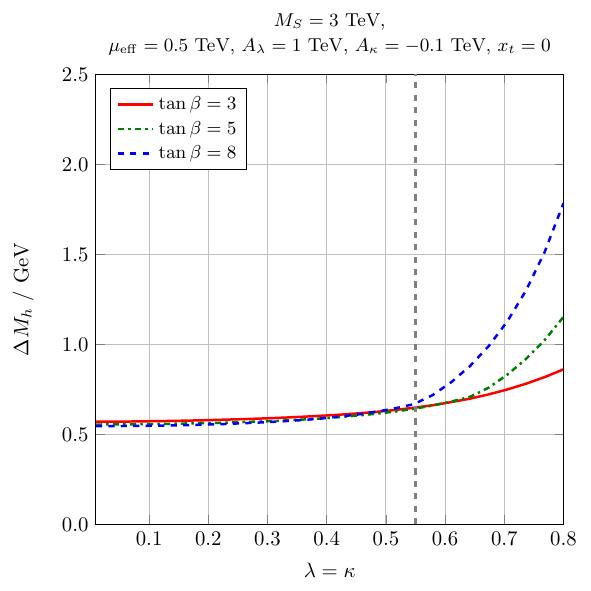}
  \end{subfigure}
  \caption{On the left side, the differences between the Higgs masses
    calculated by using the \FEFTH\ approach with $n$-loop matching depending on
    $\lambdaNMSSM$ and $\kappa$ are shown. The dependence of the estimated
    theory uncertainty $\Delta M_h$ corresponding to missing higher
    order contributions for different values of $\tan \beta$ is
    depicted on the right side. Here, the vertical dashed line
    indicates the perturbativity limit, see Eq.~\eqref{eq:Pert_Lim}.}
  \label{fig:Mh_lambda_kappa_2}
\end{figure}

The left plot in Figure~\ref{fig:Mh_lambda_kappa_2} shows the difference
between $M_h$ calculated at different matching loop orders in
dependence on $\lambdaNMSSM$ and $\kappa$, i.e.\ all precision settings were
choosen as presented in appendix~\ref{sec:app} with the corresponding value
for flag 21 for each case. Here $M_h^{n\ell}$ denotes the
Higgs pole mass calculated using $n$-loop matching. The curve
representing the difference between $M_h^{0\ell}$ and $M_h^{1\ell}$ indicates
large corrections and also a significant variation with
$\lambdaNMSSM=\kappa$. In the \MSSM-limit, one can read off that
non-\NMSSM-specific terms, originating mainly from top- and
stop-contributions, lead to a shift of about $-0.8\unit{GeV}$. That
difference considerably decreases in terms of absolute value when going to larger
values of $\lambdaNMSSM=\kappa$ for the regarded parameter point, which thus
demonstrates a strong dependence on these couplings. Considering the
difference between $M_h^{2\ell}$ and $M_h^{1\ell}$, one recognises
clearly less significant corrections below $1\unit{GeV}$, only slightly
depending on the two \NMSSM-specific couplings. Similarly, one finds
merely a mild dependence of $(M_h^{3\ell}-M_h^{2\ell})$, which however
increases in its absolute value with respect to $(M_h^{2\ell}-M_h^{1\ell})$.
As there is a sign change in the latter, this fact can be attributed to
an accidental compensation of different contributions. All in all, this behaviour
confirms the expectation of a good convergence of the perturbation series. It
should however be kept in mind that \NMSSM-specific 2-loop matching
corrections are not included and are therefore not taken into account
in the presented curve. In order to reassure that missing 2-loop and
higher order terms in $\lambdaNMSSM$ and $\kappa$ do not spoil the
precision of the calculation as suggested by the plot, a careful
uncertainty estimation is essential.

The right panel of Figure~\ref{fig:Mh_lambda_kappa_2}
indicates how the estimated uncertainty depends on $\lambdaNMSSM$
and $\kappa$. For small couplings ($\lambdaNMSSM=\kappa<0.2$), the uncertainty
remains nearly constant at $\Delta M_h \sim 0.7\unit{GeV}$. For
$\tan \beta = 3$, a slight positive deviation appears in comparison
to larger $\tan \beta$. However, the estimated uncertainty remains
below $1\unit{GeV}$ there. In the range of moderate couplings, slight changes
occur. While the uncertainty decreases with increasing $\lambdaNMSSM$ and
$\kappa$ for $\tan \beta = 3$, it increases for larger considered
values of $\tan \beta$. Beyond the perturbativity limit, the
uncertaintes grow quickly. That tendency is stronger for larger
$\tan \beta$. The plot thus indicates that for $\lambdaNMSSM$ and $\kappa$
allowed by the perturbativity limit, the uncertainties are of the
order of $\sim 0.7\unit{GeV}$ and only very slightly dependent on these
couplings. In contrast, for values above that limit, there is a strong
dependence entailing a significantly less reliable prediction of
$M_h$. Also, comparing to the dependence of the uncertainty on $x_t$
studied in Ref.~\cite{Kwasnitza:2020wli}, one comes to the conclusion
that the variation of $\lambdaNMSSM$ and $\kappa$ below the perturbativity
limit has only a minor numerical effect.

In order to substantiate the finding of an uncertainty of less
than $1\unit{GeV}$ for scenarios with moderate $|x_t|$, a scan with
$N=10^3$ sample points was performed. The parameters were varied
randomly in the following ranges:
\begin{subequations}
  \begin{align}
    1\unit{TeV} &\leq \MS \leq 10\unit{TeV}, & 2 & \leq \tan\beta \leq 20, \\
    0.01 & \leq \lambdaNMSSM, \kappa \leq 0.70, & 200\unit{GeV} & \leq \mueff \leq 2500\unit{GeV}, \\
    -2000\unit{GeV} & \leq A_{\lambdaNMSSM} \leq 5000\unit{GeV}, & -2000\unit{GeV} & \leq A_{\kappa} \leq 0\unit{GeV}, \\
    -\sqrt{6} & \leq x_t \leq \sqrt{6}.
  \end{align}%
\end{subequations}
During the scan, parameter choices leading to a wrong vacuum structure were ignored
and generated anew. Among the points produced in this way, those violating the
perturbativity limit for $\lambdaNMSSM$ and $\kappa$ were discarded. For each of the
remaining 796 points, the theory uncertainty was estimated as described
in the previous section. It was found that $\Delta M_h < 0.8\unit{GeV}$
holds for 86.2\% and $\Delta M_h < 1\unit{GeV}$ for 97.9\% of the parameter points.
The majority of parameter points with higher uncertainties $\Delta M_h > 1\unit{GeV}$ have large $|x_t|$.
Nevertheless, even for large $|x_t|$ the typical uncertainty is small: out of the
144 points with $|x_t|>2$ in the scan, 94\% have $\Delta M_h<1\unit{GeV}$ GeV. Below, we will
provide additional details on cases with large $|x_t|$. 

\subsubsection{Benchmark scenarios for \boldmath$M_h^{\text{exp}}$}

Next, some \NMSSM\ benchmark scenarios are presented which reproduce
the experimentally determined Higgs mass $M_h$ within the
1$\sigma$-uncertainty band according to the 3-loop \FEFTH\ calculation
as implemented in \FS. For the chosen points, the conditions from
Eqs.~\eqref{eq:parameter_condition_MSSM_mf}--\eqref{eq:parameter_condition_MSSM_Af}
are fulfilled again. The \SUSY\ mass scale $\MS$ is set
to $3\unit{TeV}$ for each point. All other parameter are set as in
Table~\ref{tab:BMP_Par}.

The parameter values were chosen such that small, moderate
and large values of $\tan \beta$ are covered (corresponding to the values 3, 7
and 20). In order to achieve a sufficiently high $M_h$, the
\NMSSM-specific couplings $\lambdaNMSSM$ and $\kappa$ are accordingly chosen
to be small ($\sim 0.05$) or large ($\sim 0.6$) compared to the
perturbativity limit. For each value of $\tan \beta$, one parameter
point with a large mass splitting between the Higgs bosons and one
point with a significantly smaller splitting were considered by adapting
the value $\mueff$. Finally, the remaining parameters were
adjusted such that the experimental Higgs mass value is
reproduced. Furthermore, one scenario was added with $\lambdaNMSSM$ and
$\kappa$ beyond the perturbativity limit. The chosen set of points
thus covers qualitatively very different situations that can occur
within the \NMSSM.

\begin{table}
  \centering
  \begin{tabular}{ccccccccc}
    \toprule
    BMP & $\tan \beta$ & $\lambdaNMSSM$ & $\kappa$ & $\mueff$/GeV & $A_{\lambdaNMSSM}$/GeV & $A_{\kappa}$/GeV & $x_t$ \\ \midrule
    S1  & $20$ & $0.05$ & $0.05$ & $800$ & $1000$ & $-500$  & $1.85$ \\
    S2  & $20$ & $0.05$ & $0.05$ & $300$ & $-250$ & $-950$  & $2.15$ \\
    S4  &  $7$ & $0.30$ & $0.30$ & $800$ & $5000$ & $-100$  & $2.05$ \\
    S5  &  $7$ & $0.30$ & $0.30$ & $300$ & $1200$ & $-10$   & $2.50$ \\
    S7  &  $3$ & $0.60$ & $0.30$ & $750$ & $2000$ & $-1000$ & $1.60$ \\
    S8  &  $3$ & $0.55$ & $0.35$ & $300$ & $ 700$ & $-600$  & $1.85$ \\
    S10 &  $3$ & $1.50$ & $1.50$ & $800$ & $ 250$ & $-1000$ & $0.25$ \\ \bottomrule
  \end{tabular}
  \caption{Chosen parameter values for the analyzed benchmark points.}
  \label{tab:BMP_Par}
\end{table}

In Table~\ref{tab:BMP_Spec} we show the Higgs mass spectra as obtained from
\FS\ as well as the estimation of the uncertainties connected to missing higher
order terms for the mass of the \SM-like Higgs for the selected benchmark
points. The values of the total uncertainty are each of the order
$1\unit{GeV}$, except for benchmark point S10 whose values for $\lambdaNMSSM$
and $\kappa$ are significantly beyond the perturbativity limit. This
observation indicates again that the calculation presented here
generates reliable results for various plausible scenarios within the
\NMSSM. The apparent fluctuation in the uncertainties can mostly be
attributed to the different chosen values of $\mueff$ and $x_t$. The
large uncertainty determined for S10 can however be directly ascribed
to the large values of $\lambdaNMSSM$ and $\kappa$, according to the
findings of the previous subsection.

\begin{table}
  \centering
  \begin{tabular}{crrrrrrrr}
    \toprule
    BMP & $M_{h_1}$ & $M_{h_2}$ & $M_{h_3}$ & $M_{A_1}$ & $M_{A_2}$ & $M_{H^{\pm}}$ & $\Delta M_{h_1}$ & $\Delta M_{h_1}^{\HS}$ \\ \midrule
     S1 & $125.17$ & $1469.64$ & $5346.08$ & $1095.27$ & $5346.12$ & $5347.01$ &  $0.80$ &  $0.56$ \\
     S2 & $125.11$ &  $274.45$ &  $433.03$ &  $432.72$ &  $924.38$ &  $441.98$ &  $0.60$ &  $0.36$ \\
     S4 & $124.83$ & $1574.00$ & $5744.72$ &  $476.90$ & $5745.17$ & $5745.41$ &  $0.86$ &  $0.62$ \\
     S5 & $125.16$ &  $590.79$ & $1794.10$ &  $136.82$ & $1793.29$ & $1794.25$ &  $1.03$ &  $0.79$ \\
     S7 & $125.37$ &  $451.88$ & $2421.85$ & $1070.46$ & $2420.73$ & $2418.73$ &  $0.68$ &  $0.44$ \\
     S8 & $125.33$ &  $169.45$ &  $935.33$ &  $593.01$ &  $931.83$ &  $928.66$ &  $0.74$ &  $0.50$ \\
    S10 & $125.11$ & $1159.58$ & $1612.76$ & $1358.12$ & $1625.07$ & $1572.56$ & $15.13$ & $14.89$ \\ \bottomrule
  \end{tabular}
  \caption{Calculated Higgs mass spectra for the benchmark points
    defined above. The estimated total uncertainty for the mass of the
    SM-like Higgs boson $h_1$ is also listed as well as its high-scale
    fraction. The low-scale uncertainty is nearly the same for all
    scenarios with about $0.3\unit{GeV}$, its exact value can be calculated
    using Eq.~\eqref{eq:Comb_Uncert}. All masses and uncertainties are
    given in GeV.}
  \label{tab:BMP_Spec}
\end{table}

\section{Summary and outlook}

In this paper, we presented an analysis of the \SM-like Higgs mass
in the \MSSM\ and its dependencies on various parameters, based on the
implementation of the \FEFTH\ hybrid approach in \FS. The various
scenarios regarded here are characterized by non-degenerated \SUSY\
mass spectra, complementing the consideration of ``standard''
scenarios regarded in our previous study. In general, we found a very
good reliability of the \FEFTH\ calculation for the large majority of
the analysed cases according to the estimated theoretical
uncertainty. The uncertainty is typically of the order of
$\sim 0.5\unit{GeV}$. However, specific scenarios may lead to
significantly higher uncertainties, especially ones that have
a significant splitting in the third generation squark mass
parameters.

We reported also about the most recent implementation of the \FEFTH\
calculation in the \NMSSM, now publicly avaiable as a part of
version~2.9.0 of \FS. It incorporates a treatment in the full-model
parametrization as well as the inclusion of \MSSM-like 2-loop and
3-loop contributions to the quartic \SM\ Higgs coupling. This
\NMSSM\ hybrid calculation thus provides a Higgs mass prediction at
\NNNLO\ with a resummation of large logarithms of \NNNLL\ order
in the respective couplings.  In this way, significant progress is
achieved in comparison to the previous
implementation of the \FEFTH\ method for the \NMSSM\ in \FS.
The results obtained from applying the calculation to several
different scenarios indicate a high stability of the code. In
general, the 3-loop \SUSY\ \QCD\ contributions have a rather small
impact on the Higgs mass, e.g. of the order of $0.1\unit{GeV}$
for degenerate mass spectra with $M_S \approx 3\unit{TeV}$. However,
the 3-loop calculation can suffer from large uncertainties and
discontinuities due to hierarchy transitions which might be
problematic in some applications. Therefore we recommend to use the
2-loop calculation when the 3-loop calculation leads to significantly
larger uncertainties or if continuity of the Higgs mass prediction is required.

The reliability of the Higgs pole mass calculation in the \NMSSM\ has
been further corroborated by studying the potential impact of
\NMSSM-specific missing higher-order terms. It was confirmed that
matching scale variation is an appropriate means to simulate the
effect of those disregarded contributions, unless very specific
conditions are fulfilled, like $\kappa \gg \lambdaNMSSM$. The theoretical
uncertainty was found to be nearly constant all over the
parameter space allowed by the perturbativity limit on $\lambdaNMSSM$ and
$\kappa$ with $\Delta M_h \approx 0.7\unit{GeV}$, making it one of the
most precise calculations of the SM-like Higgs boson mass in the
\NMSSM\ currently available.

\section*{Acknowledgements}
We thank the \NMSSMCalc developers and in particular M.\ Gabelmann for supporting the comparison to \FS and the clarification the differences between the two codes. This research was supported by the German Research Foundation (DFG) under grant number STO 876/2-3 and by the high-performance computing cluster Taurus at ZIH, TU Dresden.

\clearpage
\appendix
\section{SUSY Les Houches Input}
\label{sec:app}

The \SUSY\ Les Houches (\SLHA) input file for \FS's
\texttt{NMSSMEFTHiggs} model used for the benchmark point S1 is given
below. The precision settings for all other 3-loop \FEFTH\
calculations presented in this publication are chosen identical.
\begin{lstlisting}
Block FlexibleSUSY
    0   1.0e-05    # precision goal
    1   0          # max. iterations (0 = automatic)
    2   0          # algorithm (0 = all, 1 = two_scale, 2 = semi_analytic)
    3   0          # calculate SM pole masses
    4   3          # pole mass loop order
    5   3          # EWSB loop order
    6   4          # beta-functions loop order
    7   3          # threshold corrections loop order
    8   1          # Higgs 2-loop corrections O(alpha_t alpha_s)
    9   1          # Higgs 2-loop corrections O(alpha_b alpha_s)
   10   1          # Higgs 2-loop corrections O((alpha_t + alpha_b)^2)
   11   1          # Higgs 2-loop corrections O(alpha_tau^2)
   12   0          # force output
   13   2          # Top pole mass QCD corrections (0 = 1L, 1 = 2L, 2 = 3L, 3 = 4L)
   14   1.0e-11    # beta-function zero threshold
   15   0          # calculate all observables
   16   0          # force positive majorana masses
   17   0          # pole mass renormalization scale (0 = SUSY scale)
   18   0          # pole mass renormalization scale in the EFT (0 = min(SUSY scale, Mt))
   19   0          # EFT matching scale (0 = SUSY scale)
   20   2          # EFT loop order for upwards matching
   21   3          # EFT loop order for downwards matching
   22   0          # EFT index of SM-like Higgs in the BSM model
   23   1          # calculate BSM pole masses
   24   124111321  # individual threshold correction loop orders
   25   0          # ren. scheme for Higgs 3L corrections (0 = DR, 1 = MDR)
   26   1          # Higgs 3-loop corrections O(alpha_t alpha_s^2)
   27   0          # Higgs 3-loop corrections O(alpha_b alpha_s^2)
   28   0          # Higgs 3-loop corrections O(alpha_t^2 alpha_s)
   29   0          # Higgs 3-loop corrections O(alpha_t^3)
   30   0          # Higgs 4-loop corrections O(alpha_t alpha_s^3)
   31   0          # loop library (0 = softsusy)
Block SMINPUTS            # Standard Model inputs
    1   1.279160000e+02   # alpha^(-1) SM MSbar(MZ)
    2   1.166378700e-05   # G_Fermi
    3   1.184000000e-01   # alpha_s(MZ) SM MSbar
    4   9.118760000e+01   # MZ(pole)
    5   4.18000000e+00    # mb(mb) SM MSbar
    6   1.73340000e+02    # mtop(pole)
    7   1.77700000e+00    # mtau(pole)
    8   0.00000000e+00    # mnu3(pole)
    9   80.384            # MW pole
   11   5.10998902e-04    # melectron(pole)
   12   0.00000000e+00    # mnu1(pole)
   13   1.05658357e-01    # mmuon(pole)
   14   0.00000000e+00    # mnu2(pole)
   21   4.75000000e-03    # md(2 GeV) MS-bar
   22   2.40000000e-03    # mu(2 GeV) MS-bar
   23   1.04000000e-01    # ms(2 GeV) MS-bar
   24   1.27000000e+00    # mc(mc) MS-bar
Block EXTPAR
   0   3000   # MSUSY
   1   3000   # M1
   2   3000   # M2
   3   3000   # M3
   4   3000   # Mu
   25  20     # TanBeta
   61  0.05   # Lambda
   62  0.05   # Kappa
   63  1000   # ALambda
   64 -500    # AKappa
Block MSQ2IN
  1  1     9.0e+06   # mq2(1,1)
  2  2     9.0e+06   # mq2(2,2)
  3  3     9.0e+06   # mq2(3,3)
Block MSE2IN
  1  1     9.0e+06   # me2(1,1)
  2  2     9.0e+06   # me2(2,2)
  3  3     9.0e+06   # me2(3,3)
Block MSL2IN
  1  1     9.0e+06   # ml2(1,1)
  2  2     9.0e+06   # ml2(2,2)
  3  3     9.0e+06   # ml2(3,3)
Block MSU2IN
  1  1     9.0e+06   # mu2(1,1)
  2  2     9.0e+06   # mu2(2,2)
  3  3     9.0e+06   # mu2(3,3)
Block MSD2IN
  1  1     9.0e+06   # md2(1,1)
  2  2     9.0e+06   # md2(2,2)
  3  3     9.0e+06   # md2(3,3)
Block AUIN
  1  1     0   # Au(1,1)
  2  2     0   # Au(2,2)
  3  3  5590   # Au(3,3)
Block ADIN
  1  1     0   # Ad(1,1)
  2  2     0   # Ad(2,2)
  3  3     0   # Ad(3,3)
Block AEIN
  1  1     0   # Ad(1,1)
  2  2     0   # Ad(2,2)
  3  3     0   # Ad(3,3)
\end{lstlisting}
For the 2-loop \FEFTH\ calculations performed in this work the flag~21 in the \FS\ block is set to~2 and analogously for the 1-loop calculations. 
\clearpage
\printbibliography[heading=bibintoc]
\end{document}